\documentclass[cernpreprint, atlasdraft=false, UKenglish, texmf, orcidlogo]{atlasdoc}

\usepackage[subfigure]{atlaspackage}
\usepackage{atlasbiblatex}
 
\usepackage{atlasphysics}
\usepackage{xcolor}
\usepackage{graphicx}
\usepackage{float}
\usepackage{xcolor}
\usepackage{longtable}
\usepackage{amsmath}
\usepackage{numprint}
\usepackage{fixltx2e}
\usepackage{comment}
\usepackage{multirow}
\usepackage{nicematrix}
\usepackage{comment}
\usepackage{lineno}

\graphicspath{{logos/}{figures/}}

\newcommand{\Prompt}{\ensuremath{\mathrm{P}}}
\newcommand{\NPrompt}{\ensuremath{\mathrm{NP}}}

\AtlasTitle{Measurement of the production cross-section of $J/\psi$ and
$\psi(2\mathrm{S})$ mesons in $pp$ collisions at
$\sqrt{s} = 13$~TeV with the ATLAS detector
}

\AtlasAbstract{
Measurements of the differential production cross-sections of prompt and non-prompt $J/\psi$ and $\psi(2\mathrm{S})$
mesons with transverse momenta
between 8 and 360~GeV and rapidity in the range $|y|<2$ are reported.
Furthermore, measurements of the non-prompt fractions of $J/\psi$ and $\psi(2\mathrm{S})$,
and the prompt and non-prompt $\psi(2\mathrm{S})$-to-$J/\psi$ production ratios, are presented.
The analysis is performed using 140~fb$^{-1}$ of $\sqrt{s}=13$~TeV $pp$ collision data recorded by the ATLAS detector at the LHC
during the years 2015--2018.}
 
\AtlasRefCode{BPHY-2019-08}
 
\PreprintIdNumber{CERN-EP-2023-193}

\AtlasJournal{Eur. Phys. J. C}
\AtlasJournalRef{Eur. Phys. J. C 84 (2024) 169}
\AtlasDOI{10.1140/epjc/s10052-024-12439-9}

\hypersetup{pdftitle={ATLAS document},pdfauthor={The ATLAS Collaboration}}

\begin{document}
 
\maketitle

\section{Introduction}
\label{sec:intro}

% The next lines are included from the .//Introduction.tex input file
Studies involving heavy quarkonia
provide a unique insight into the nature of quantum
chromodynamics (QCD) near the boundary of the perturbative and non-perturbative regimes.
However, despite a long history of research, quarkonium production in hadronic collisions
still presents significant challenges to both theory and experiment.
 
In high-energy hadronic collisions, charmonium states can be produced either from
short-lived QCD sources (referred to as \textit{prompt} production) or from long-lived sources -- decays
of beauty hadrons (\textit{non-prompt} production).
These two sources can be distinguished experimentally by measuring the distance between
the production and decay vertices of the charmonium state.
Feed-down decays of higher charmonium states contribute to the production of $\jpsi$ mesons for both of these sources, and should be taken into account when comparing with theoretical predictions
(there is no significant feed-down contribution to $\psip$ production).
While calculations within the framework of perturbative QCD (see e.g.\ Refs.~\cite{FONLL_2001,FONLL_2012})
have been reasonably successful in describing the non-prompt contributions,
a satisfactory understanding of the prompt production mechanisms is still
to be achieved.

Methods developed within the non-relativistic QCD (NRQCD) approach provide a framework for describing quarkonium production processes,
leading to a variety of models differing in their accuracy and predictive power.
In particular, Ref.~\cite{Bodwin:1994jh} introduced a number of phenomenological parameters -- long-distance matrix elements (LDMEs) -- which can be extracted from fits to the experimental data, and are expected to
describe the cross-sections and differential spectra for several sets of data reasonably well~{\cite{BPHY-2012-02,BPHY-2013-06,Aaij:2012ag,Chatrchyan:2011kc}}.
However, various attempts to build a universal library of LDMEs to be used to describe a wider range of measurements such as the polarisation of quarkonia~\cite{Abelev:2011md,CMS-BPH-13-003,Aaij:2014qea,Acharya:2018uww}, their associated
production~\cite{Aad:2014rua,Aad:2014kba}  or the production of quarkonium in a wider range of processes (e.g.\ photo- and electro-production)
have not been particularly
successful~\cite{Li:2010hc,Lansberg:2013wva,Gong:2012ah, Mao:2011kf,Butenschoen:2012qh}.
An alternative approach to the description of the hadronisation process of heavy quarkonia is offered by
the Colour Evaporation Model (CEM)~\cite{Barger:1979js,Barger:1980mg,Gavai:1994in}, which offers a simpler framework with fewer parameters, but has its own problems in describing the data~\cite{Ma:2016exq}.
A combination of ATLAS results with cross-section and polarisation measurements
from CMS \cite{Chatrchyan:2011kc,CMS-BPH-14-001,CMS-BPH-15-005},
LHCb \cite{Aaij:2012ag,LHCb2,LHCb3,LHCb4,LHCb5,LHCb6}  and ALICE
\cite{Abelev:2014qha,ALICE2021,ALICE1,ALICE2,Acharya:2018uww,ALICE4,ALICE5,ALICE:2011gej} now includes a variety of charmonium production characteristics in a wide
kinematic range, thus providing a wealth of information
for a new generation of theoretical models.

One way to add qualitatively new information is to extend the kinematic range of quarkonium production measurements.
ATLAS has previously measured the inclusive differential cross-section for $\jpsi$
production in $pp$ collisions at $\sqrt{s} = 7$ and $8$~\TeV~\cite{BPHY-2012-02},
as well as the differential cross-sections for the production
of $\chi_c$ states~\cite{BPHY-2013-05}, and for $\psip$ production~\cite{BPHY-2013-06}.
In most of these measurements, ATLAS exploited a dimuon trigger with a muon transverse momentum ($\pt$) threshold of 4~\si{\GeV}, with the high-$\pt$ reach
limited mainly by the dimuon trigger's performance to about 100~\si{\GeV}: at higher $\pt$ values the angular resolution of the muon trigger system is not sufficient to separate the two almost collinear muons.
This paper describes a measurement of
$\jpsi$ ($\psip$) meson production via decay in the dimuon channel, at $\sqrt{s} = 13$~\TeV\ for meson transverse momenta of 8--360~\GeV\ (8--140~\GeV), which is a much broader range than in previous measurements.
This was made possible by the use of two different triggers. Production of $\jpsi$ and $\psip$ at low $p_{\mathrm{T}}$, between 8 and $60~\si{\gev}$, is measured using a dimuon trigger requiring a pair of muons to each pass a $p_{\mathrm{T}}$ threshold of $4~\si{\gev}$,
while at high $p_{\mathrm{T}}$ a single-muon trigger with a $\pt$ threshold of $50~\si{\gev}$ was used.
This allowed measurements to be performed for transverse momenta as high as
$360~\si{\gev}$  for $\jpsi$ and $140~\si{\gev}$ for $\psip$.
The measurements include the double-differential cross-sections for production of the two vector charmonium states (separately for the prompt
and non-prompt production mechanisms), the non-prompt fraction for each state, and the
prompt and non-prompt $\psip$-to-$\jpsi$ production ratios.

The paper is organised as follows. A brief description of the ATLAS detector is given in Section~\ref{sec:detector}. The event selection and the analysis strategy are explained in Section~\ref{sec:strategy}, followed by a description of the systematic uncertainties affecting the measurement in Section~\ref{sec:systematics}.
Results and comparisons with theoretical calculations are presented in Section~\ref{sec:results}, followed by a summary in Section~\ref{sec:conclusion}.

% End of text imported from the .//Introduction.tex input file

\section{The ATLAS detector}
\label{sec:detector}

% The next lines are included from the .//Detector.tex input file
\newcommand{\AtlasCoordFootnote}{
ATLAS uses a right-handed coordinate system with its origin at the nominal interaction point (IP)
in the centre of the detector and the \(z\)-axis along the beam pipe.
The \(x\)-axis points from the IP to the centre of the LHC ring,
and the \(y\)-axis points upwards.
Cylindrical coordinates \((r,\phi)\) are used in the transverse plane,
\(\phi\) being the azimuthal angle around the \(z\)-axis.
The pseudorapidity is defined in terms of the polar angle \(\theta\) as \(\eta = -\ln \tan(\theta/2)\).
Angular distance is measured in units of \(\Delta R \equiv \sqrt{(\Delta\eta)^{2} + (\Delta\phi)^{2}}\).}

The ATLAS experiment~\cite{PERF-2007-01} at the LHC is a multipurpose particle detector
with a forward--backward symmetric cylindrical geometry and a near \(4\pi\) coverage in
solid angle.\footnote{\AtlasCoordFootnote}
It consists of an inner tracking detector (ID) surrounded by a thin superconducting solenoid
providing a \qty{2}{\tesla} axial magnetic field, electromagnetic and hadron calorimeters, and a muon spectrometer.
The inner tracking detector covers the pseudorapidity range \(|\eta| < 2.5\).
It consists of silicon pixel, silicon microstrip, and transition radiation tracking detectors.
Metal/liquid-argon (LAr) sampling calorimeters provide electromagnetic (EM) energy measurements
with high granularity.
A steel/scintillator-tile hadron calorimeter covers the central pseudorapidity range (\(|\eta| < 1.7\)).
The endcap and forward regions are instrumented with LAr calorimeters
for both the EM and hadronic energy measurements up to \(|\eta| = 4.9\).
The muon spectrometer surrounds the calorimeters and is based on
three large superconducting air-core toroidal magnets with eight coils each.
The field integral of the toroids ranges between \num{2.0} and \qty{6.0}{\tesla\metre}
across most of the detector.
The muon spectrometer has a system of precision chambers for tracking and fast detectors for triggering.
A two-level trigger system is used to select events.
The first-level trigger is implemented in hardware and uses a subset of the detector information
to accept events at a rate below \qty{100}{\kHz}.
This is followed by a software-based trigger that
reduces the accepted event rate to \qty{1}{\kHz} on average
depending on the data-taking conditions.
An extensive software suite~\cite{ATL-SOFT-PUB-2021-001} is used in data simulation, in the reconstruction
and analysis of real and simulated data, in detector operations, and in the trigger and data acquisition
systems of the experiment.

% End of text imported from the .//Detector.tex input file

\section{Analysis strategy}
\label{sec:strategy}

% The next lines are included from the .//AnalysisStrategy.tex input file
\subsection{Event selection}
\label{ssec:eventsel}
 
Data for this analysis were taken during the LHC proton--proton collision runs at
$\sqs = 13~\si{\tev}$ in the years 2015 to 2018.
For lower values of the transverse momentum $\pt$ of the dimuon system, between 8 and $60~\si{\gev}$,
a dimuon trigger was used, requiring a pair of muons to each pass a $p_{\mathrm{T}}$ threshold of $4~\si{\gev}$. This trigger ran unprescaled during 2015 data-taking, collecting an integrated luminosity of  2.6~fb$^{-1}$. In the high $p_{\mathrm{T}}$ range between 60 and $360~\si{\gev}$, a single-muon trigger with a $\pt$ threshold of $50~\si{\gev}$ was used, unprescaled throughout the full \RunTwo data-taking, providing a total integrated luminosity
of 140~\si{\per\fb}.
The selected events were required to contain a pair of oppositely charged muons of high quality (using the {\textit{tight}} identification requirements defined in Ref.~\cite{PERF-2015-10}),
with $\pt>4~\si{\GeV}$ and $|\eta|<2.4$.
In the low $\pt$ range the two muons were required to match the two trigger objects of the dimuon trigger, while in the high $\pt$ range at least one of the muons was required to have
$\pt > 52.5~\si{\GeV}$ and match the
trigger object.
The two ID tracks attributed to the muons were fitted to a common vertex,
and the dimuon invariant mass $m_{\mu\mu}$ was required to satisfy
$2.6 < m_{\mu\mu} < 4.2~\si{\GeV}$.
The transverse distance $L_{xy}$ between the primary vertex and the dimuon vertex was used to calculate the meson's pseudo-proper decay time
\begin{equation*}
\tau = \frac{m_{\mu\mu}}{\pt}\frac{L_{xy}}{c},
\end{equation*}
where $\pt$ is the reconstructed transverse momentum of the dimuon system, and $c$ is the speed of light.
The primary vertex is chosen as the reconstructed collision interaction vertex whose $z$ coordinate is nearest to the point of closest approach of the dimuon system's trajectory to the beam axis.
If an event has more than one selected dimuon candidate, all candidates are retained and treated independently.
 
\subsection{Cross-section determination}
\label{ssec:crossec}
 
The phase space of the measurement is divided into 34 intervals in
dimuon $\pt$ covering the range from 8 to 360~\si{\GeV}, and 3 intervals in absolute rapidity{\footnote{Rapidity $y$ of a particle of energy $E$ and longitudinal momentum $p_z$ is defined as $y=\frac{1}{2}\log\frac{E+p_z}{E-p_z}$.}}
$|y|$ with boundaries at 0, 0.75, 1.5 and 2.0, thus producing 102 analysis bins overall.
In each $(p_{\mathrm{T}},y)$ 
bin, a two-dimensional unbinned maximum-likelihood fit to the distribution of dimuon candidates in invariant mass $m_{\mu\mu}$ and pseudo-proper decay time $\tau$ of the $\psi$ meson is performed to obtain the raw yields $N_{\psi}^\text{P,NP}$ for prompt (P) and non-prompt (NP) $\psi$ mesons, where $\psi = J/\psi, \; \psi(2S)$.
The raw yields are then corrected to account for the geometrical acceptance ${\mathcal{A}(\psi)}$,  the trigger and reconstruction efficiencies $\epsilon_{\mathrm{trig}}$ and
$\epsilon_{\mathrm{reco}}$, and the trigger and reconstruction correction scale factors $ \epsilon_{\mathrm{trigSF}}$ and $ \epsilon_{\mathrm{recoSF}}$,
averaged over that 
bin.
Several low $\pt$ bins are divided into narrower sub-bins to obtain raw yields in finer granularity, which are then corrected and summed to give the final yield in the corresponding analysis bin. This procedure helps to reduce measurement biases due to modelling assumptions in the regions of phase space with large statistical power.
 
The prompt (P) and non-prompt (NP) double-differential production cross-sections for $\psi = J/\psi, \; \psi(2S)$ in each analysis bin are calculated as
\begin{equation}
\frac{\dif^{2}\sigma^\text{P,NP} (pp\rightarrow \psi )}{\dif p_{\mathrm{T}}\dif y}\times \mathcal{B}(\psi \rightarrow \mu^+ \mu^-)
= \frac{1}{{\mathcal{A}(\psi)} \epsilon_{\mathrm{trig}} \epsilon_{\mathrm{trigSF}} \epsilon_{\mathrm{reco}} \epsilon_{\mathrm{recoSF}}}\;
\frac{N_{\psi}^\text{P,NP}}{\;\Delta p_{\mathrm{T}}\; \Delta y \;\int\!\mathcal{L} \dif t},
\label{eq:cross_section}
\end{equation}
where $\Delta p_{\mathrm{T}}$ and $\Delta y$ are bin widths in $p_{\mathrm{T}}$ and rapidity, and $\int\!\mathcal{L} \dif t$ is the corresponding integrated luminosity. Bin migration effects are discussed in Section~\ref{sec:eff_corr}.
The acceptance ${\mathcal{A}}(\psi)$ is defined as the probability that a $\psi$ state with
(true) momentum within an analysis bin survives the following acceptance selections imposed on the two muons (assuming $\pt(\mu1)>\pt(\mu2)$) in the two $\psi$ $\pt$ ranges:
\begin{itemize}
\item
low $\pt$ range, $\pt(\psi)< 60$~\GeV: $\pt(\mu_1)>4$~\GeV, $\pt(\mu_2)>4$~\GeV,
$|\eta(\mu_1)|, |\eta(\mu_2)|<2.4$;
\item
high $\pt$ range, $\pt(\psi)\ge 60$~\GeV: $p_{\mathrm{T}}(\mu_1)>52.5$~\GeV, $p_{\mathrm{T}}(\mu_2)>4$~\GeV,
$|\eta(\mu_1)|, |\eta(\mu_2)|<2.4$.
\end{itemize}
The acceptance calculation is performed using Monte Carlo (MC) generator-level kinematic variables,
with resolution effects taken into account at the efficiency correction stage.
An isotropic angular distribution of muons in the $\psi$ decay frame is assumed.
Since the spin alignment of the $\psi$ states may affect the acceptance,
a number of non-isotropic spin-alignment scenarios are used to calculate correction factors for the measured cross-sections
(see Appendix).
For a given spin-alignment scenario, systematic uncertainties due to the acceptance calculation are small (see Section~\ref{sec:systematics}).
Changing to a different spin-alignment scenario, however, can lead to noticeable changes in the cross-sections and other measured quantities,
including variations as a function of $\pt$ (see Section~\ref{sec:results}).
 
The reconstruction efficiency
$\epsilon_{\mathrm{reco}}$ and
trigger efficiency $\epsilon_{\mathrm{trig}}$ are calculated using samples of fully simulated $\jpsi$ and $\psip$ events, including appropriate trigger information.
Correction scale factors $\epsilon_{\mathrm{recoSF}}$ and $\epsilon_{\mathrm{trigSF}}$ account for the differences between simulated and real data. See Section~\ref{sec:eff_corr} for more details.
 
The non-prompt fractions for $\psi= \jpsi$, $\psip$ are defined as
\begin{eqnarray}
\centering
F^{{\NPrompt}}_{\psi}(\pt, y) &=&
\frac{\dif^{2}\sigma^{\NPrompt} (pp\rightarrow \psi )}{\dif p_{\mathrm{T}}\dif y}\times\left[\frac{\dif^{2}\sigma^{\Prompt} (pp\rightarrow \psi)}{\dif p_{\mathrm{T}}\dif y} + \frac{\dif^{2}\sigma^{\NPrompt} (pp\rightarrow \psi )}{\dif p_{\mathrm{T}}\dif y}\right]^{-1}.
\label{eq:np_fraction}
\end{eqnarray}
Finally, $\psip$-to-$\jpsi$ production ratios are defined separately for the prompt and non-prompt production mechanisms as
\begin{eqnarray}
\centering
R^{{\Prompt,\NPrompt}}(\pt, y) &=& \frac{\dif^{2}\sigma^{\Prompt, \NPrompt} (pp\rightarrow \psip )}{\dif p_{\mathrm{T}}\dif y}\times\mathcal{B}(\psip \rightarrow \mu^+ \mu^-)
\nonumber\\
&&\times\left[
\frac{\dif^{2}\sigma^{\Prompt, \NPrompt} (pp\rightarrow \jpsi )}{\dif p_{\mathrm{T}}\dif y}\times\mathcal{B}(\jpsi \rightarrow \mu^+ \mu^-)\right]^{-1}.
\label{eq:prod_ratio}
\end{eqnarray}
 
In calculating these quantities, the event yields, efficiencies, and acceptance corrections are used in accord with Eq.~(\ref{eq:cross_section}); uncertainties in the fraction and ratio measurements partially cancel out.

\subsection{Fit model}
\label{ssec:fit}
 
The fit model's probability distribution function $F(m_{\mu\mu},\tau)$ contains seven terms,
\begin{equation*}
F(m_{\mu\mu},\tau) = \sum_{i=1}^{7} \kappa_i P_i(m_{\mu\mu},\tau),
\label{eq:model0}
\end{equation*}
with fractions $\kappa_i$, describing four signal contributions and three types of background.
Terms $i=1,2$ describe prompt and non-prompt $\jpsi$ signal
respectively; terms $i=3,4$ correspond to prompt and non-prompt $\psip$ signal.
Term $i=5$ describes the prompt background, where non-resonant dimuons are produced at the primary vertex (e.g.\ Drell--Yan pairs). Term $i=6$ describes single-sided non-prompt background, mainly for dimuon continuum events where the two muons originate from the (cascade) decay of a single $b$-hadron, while term $i=7$ describes the double-sided part of the non-prompt continuum, where the two muons originate from different $b$-hadrons, yielding a secondary vertex which may appear on either side of the beamline.
For $i=2$--$7$, each term is factorised into a function $f_i(m_{\mu\mu})$ of dimuon mass $m_{\mu\mu}$ and a function $h_i(\tau)$ of pseudo-proper decay time $\tau$,
where the latter is convolved with a decay time resolution function $R(\tau)$:
\begin{equation*}
P_i(m_{\mu\mu},\tau) = f_{i}(m_{\mu\mu})\cdot \left[h_{i}(\tau)\otimes R(\tau)\right].
\label{eq:model1}
\end{equation*}
The term with $i=1$, which describes the prompt $\jpsi$ peak, has a similar structure but allows for some correlations between $m_{\mu\mu}$ and $\tau$,
as described below.

The decay time resolution function $R(\tau)$ is
parameterised as a sum of three Gaussian functions, $G_\text{A}, G_\text{B}$ and $G_\text{C}$, with the relative weights of the first two, $\omega_\text{A}$ and $\omega_\text{B}$, treated as free parameters:
\begin{equation*}
R(\tau) = \omega_\text{A} G_\text{A}(\tau) + \omega_\text{B} G_\text{B}(\tau) + (1-\omega_\text{A}-\omega_\text{B}) G_\text{C}(\tau).
\label{eq:Resol}
\end{equation*}
Based on MC studies with fully simulated signal samples, the means of the three Gaussian functions are fixed to zero, and the widths are linked by $\sigma_\text{B} = 2\sigma_\text{A}$ and $\sigma_\text{C} = 4\sigma_\text{A}$, where $\sigma_\text{A}=0.04$~ps is fixed to the smallest value found in test fits.

The parameterisations of the functions $f_i(m_{\mu\mu})$ and $h_i(\tau)$ are summarised in Table~\ref{tab:fitmod}.
The mass lineshapes  of the $\jpsi$ and $\psip$ peaks, $f_{i}(m_{\mu\mu})$ for $i=1$ to $4$, are parameterised as weighted sums of two Gaussian functions
and a Crystal Ball function~\cite{CB},
which are the same for the prompt and non-prompt components.
Based on MC studies,
the weights are common to $\jpsi$ and $\psip$, while the
ratios of the peak positions and the widths are fixed to the ratio $\beta$ of the masses~\cite{Zyla:2020zbs} of the two states. Parameters of the Crystal Ball function were kept the same for both prompt and non-prompt $\jpsi$ and also $\psip$, as verified from the MC studies.
The lifetime distributions $h_{1}(\tau)$ and $h_{3}(\tau)$ of prompt $\jpsi$ and $\psip$, respectively, are parameterised as
delta functions, while for non-prompt $\psip$, a single-sided exponential function is used for $h_{4}(\tau)$.
Since the non-prompt $\jpsi$ sample is larger and has a wider observed $\tau$ range, its decay time distribution $h_{2}(\tau)$
is described by a superposition of two single-sided exponential functions with slopes related by $\gamma_1 = b\gamma_2$,
where the constant $b=1.4$ is obtained from test fits using real data.  All these exponential functions are convolved with the resolution function $R(\tau)$.
 
In the $i=1$ term, describing the prompt $\jpsi$ peak,
the product of the narrowest Gaussian term in $f_1(m_{\mu\mu})$ and the narrowest Gaussian term in $R(\tau)$
was replaced by a bivariate Gaussian function in $m_{\mu\mu}$ and $\tau$ with a correlation coefficient $\rho=0.3$,
\begin{equation}
\omega_0 G_1(m_{\mu\mu}; \sigma_1) \cdot f_\text{A} G_\text{A}(\tau; \sigma_\text{A}) \mapsto \omega_0 f_\text{A} G_\text{BV}(m_{\mu\mu},\tau; \sigma_1, \sigma_\text{A}, \rho),
\label{eq:model2}
\end{equation}
to take into account the observed correlation between the measured values of these quantities. The effect of this correlation was found to be negligible for other terms.
 
Parameterisations for the background terms are selected using both the experience gained from similar analyses at lower energies~\cite{BPHY-2012-02} and physics considerations. In the prompt background term, $i=5$, the mass distribution is modelled by a  second-order polynomial, while the non-prompt mass distributions for $i=6$ and $7$ are
parameterised as exponential functions, with independent parameters.
The decay time distribution is a delta function for the prompt term $i=5$, a
single-sided exponential function for the main non-prompt term $i=6$, and a symmetric double-sided
exponential function for the last term, $i=7$. Each of these is also convolved with the decay time resolution function $R(\tau)$.

\begin{table}[htb]
\centering
\caption{Parameterisation of the fit
model. Here $G, CB, E$  and $P$ denote Gaussian, Crystal Ball, exponential and second-order polynomial functions, respectively, with different indices corresponding to different parameters. The parameterisation of the $i=1$ term is modified as described in the text and shown in Eq.~(\protect\ref{eq:model2}).}
\label{tab:fitmod}
\footnotesize
\begin{tabular}{c c c c c}
\hline\hline
\smallskip
i & Type & P/NP & $f_{i}(m)$ & $h_{i}(\tau)$ \\ [0.5ex]
\hline
\hline
\smallskip
1 & $J/\psi$ & P & $\omega_0 G_1(m)+(1-\omega_0)[\omega_{1}CB(m) + (1-\omega_{1})G_2(m)]$ & $\delta(\tau)$ \\  \smallskip
2 & $J/\psi$ & NP & $\omega_0 G_1(m)+(1-\omega_0)[\omega_{1}CB(m) + (1-\omega_{1})G_2(m)]$ & $\omega_2 E_1(\tau) + (1-\omega_2)E_1(b\tau)$  \\ \smallskip
3 & $\psi(2S)$ & P & $\omega_0 G_1(\beta m)+(1-\omega_0)[\omega_{1}CB(\beta m) + (1-\omega_{1})G_2(\beta m)]$ & $\delta(\tau)$ \\  \smallskip
4 & $\psi(2S)$ & NP & $\omega_0 G_1(\beta m)+(1-\omega_0)[\omega_{1}CB(\beta m) + (1-\omega_{1})G_2(\beta m)]$   &  $E_2(\tau) $ \\
\hline
5 & Bkg & P & $P$ & $\delta(\tau)$      \\ \smallskip
6 & Bkg & NP & $E_3(m)$ & $E_4(\tau)$   \\ \smallskip
7 & Bkg & NP & $E_5(m)$ & $E_6(|\tau|)$ \\[0.5ex]
\hline
\hline
\end{tabular}
\end{table}
 
Fits are performed in each (sub-)bin
using an unbinned maximum-likelihood method, in the dimuon mass range from 2.7 to 4.1~\si{\GeV} and the decay time range between $-1$ and $11$~ps.
Twenty of the 29 parameters are determined from the fit, with the rest fixed to predetermined values obtained from test fits using samples of MC simulated $\jpsi$ and $\psip$ signal events.
Uncertainties due to variations of the fit model and assumptions about the fixed parameters are estimated during the studies of systematic uncertainties described in Section~\ref{sec:systematics}.
In particular, it is found that a reliable determination of the cross-sections for prompt and non-prompt production of the $\psip$ meson
(with significance better than $5\sigma$) is only possible up to $\pt = 140$~\si\GeV, mainly due to poorer mass resolution and a lower signal-to-background ratio at higher $\pt$.
For $\pt>140$~\si\GeV\ the yield of $\psip$ was fixed to a constant fraction (0.07) of the yield of $\jpsi$, by extrapolating the results from the fits at lower transverse momenta.

\begin{figure}[H]
\centering
\subfigure[]{\includegraphics[width=.43\columnwidth]{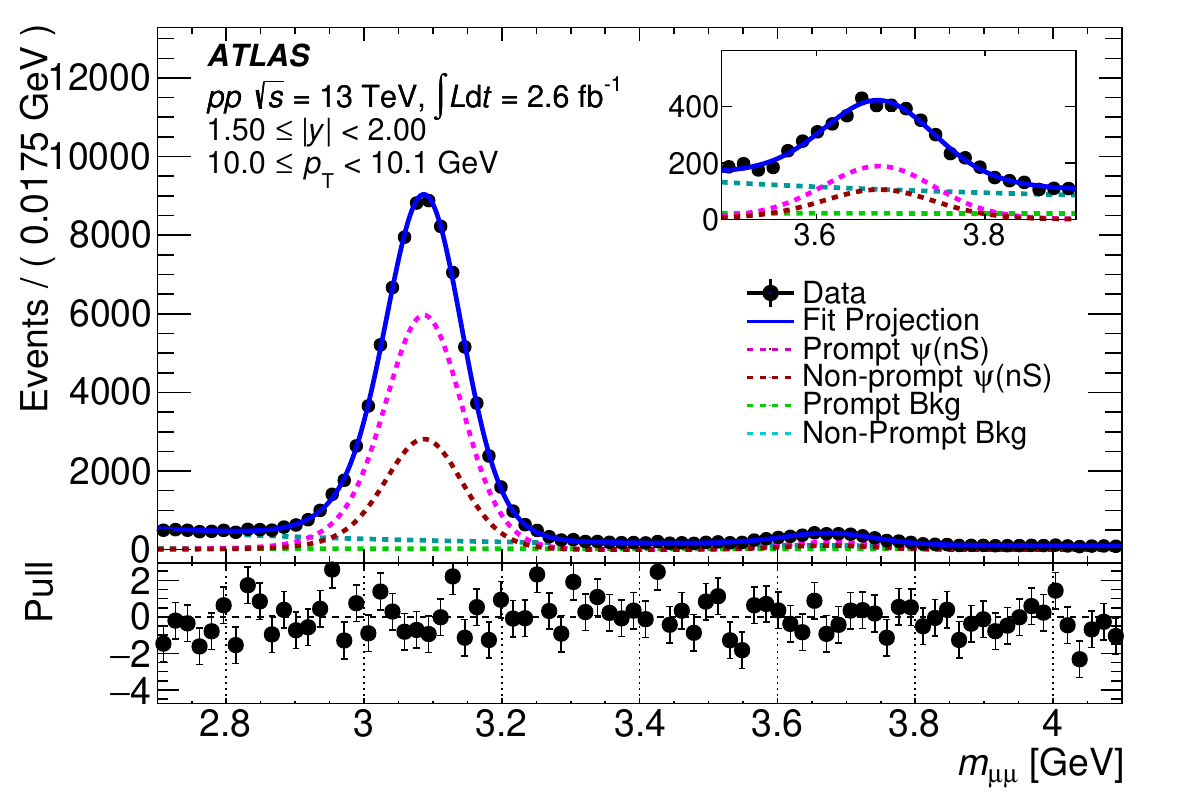}}
\subfigure[]{\includegraphics[width=.43\columnwidth]{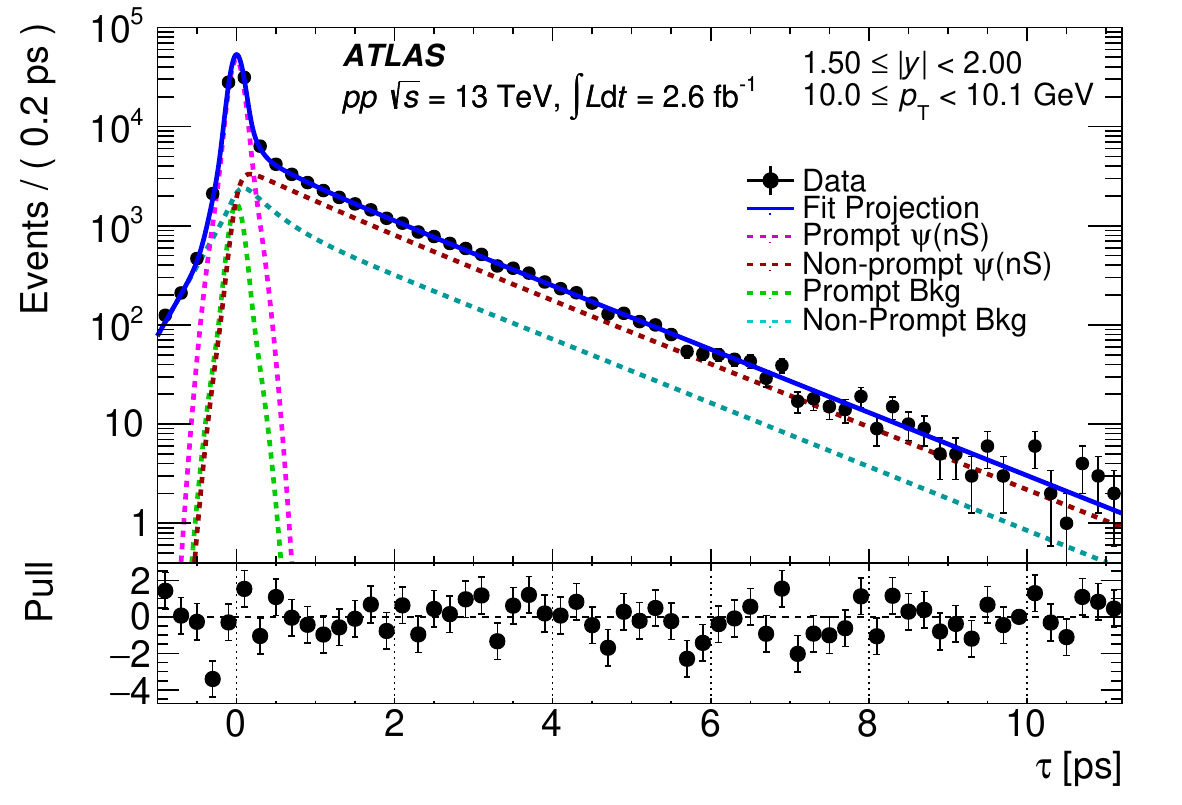}}
\subfigure[]{\includegraphics[width=.43\columnwidth]{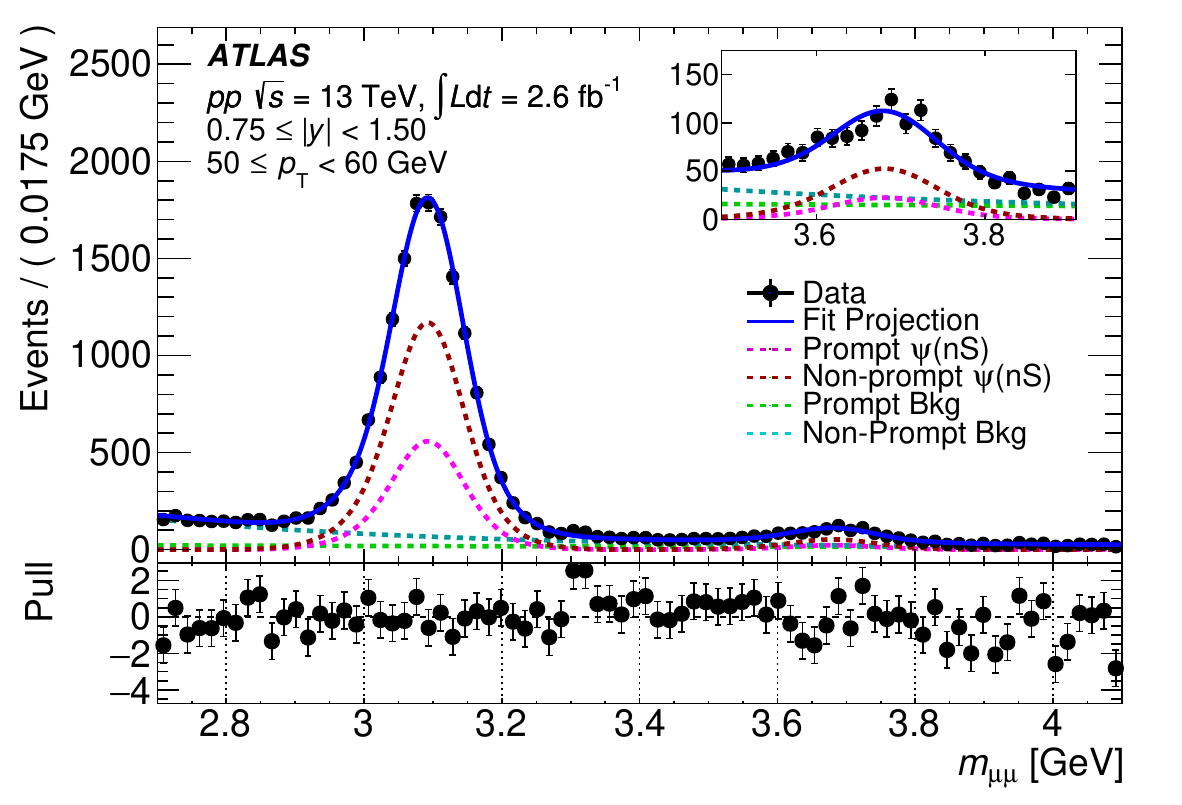}}
\subfigure[]{\includegraphics[width=.43\columnwidth]{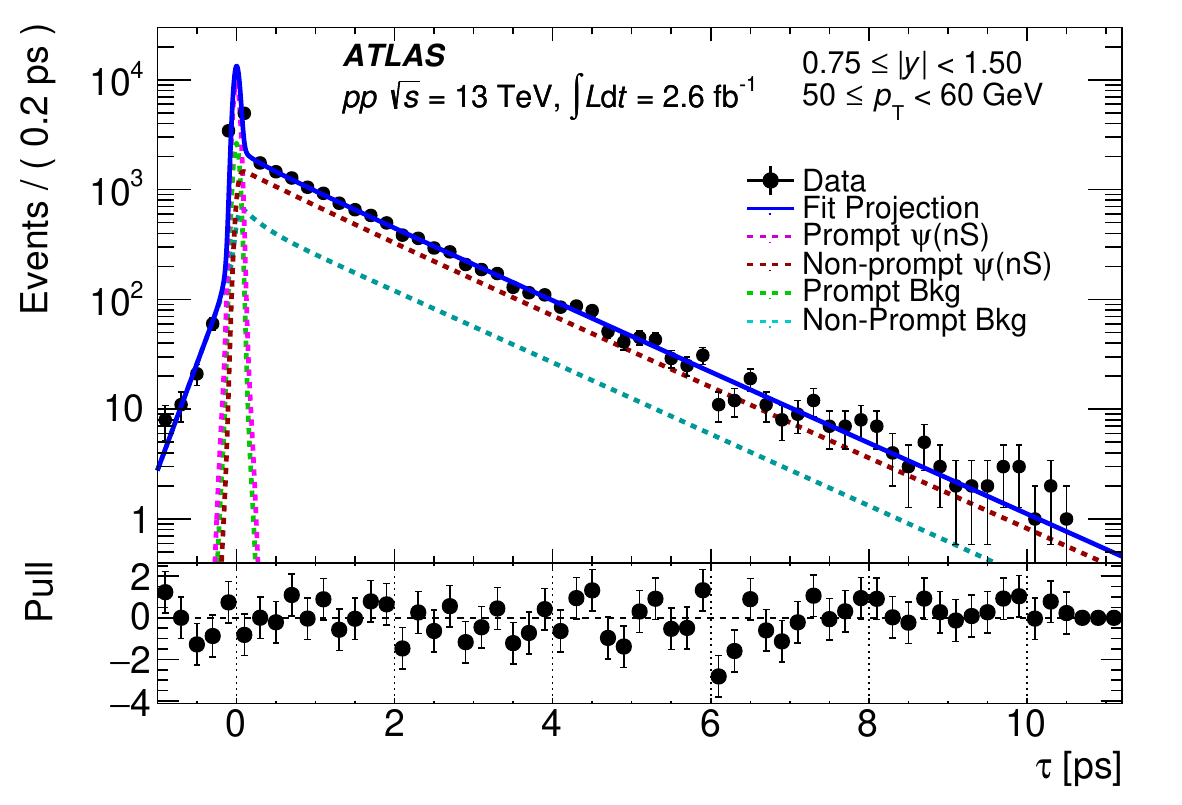}}
 
\subfigure[]{\includegraphics[width=.43\columnwidth]{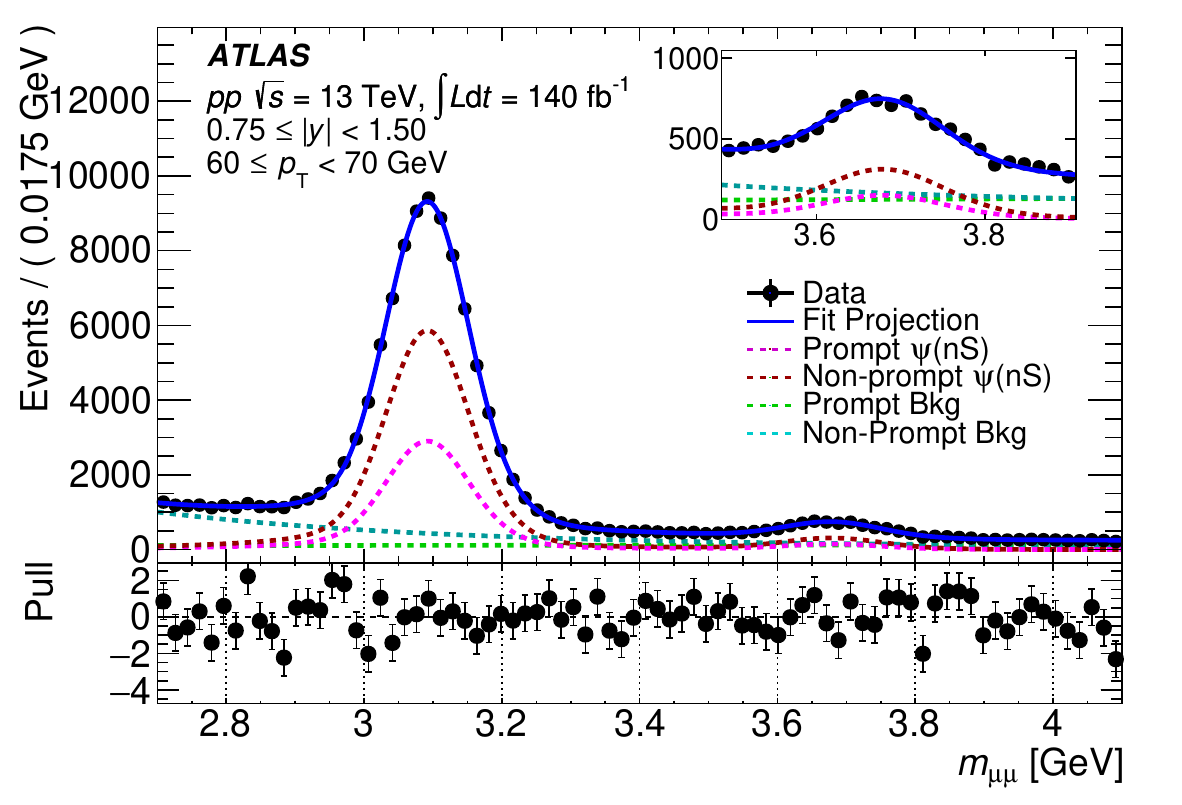}}
\subfigure[]{\includegraphics[width=.43\columnwidth]{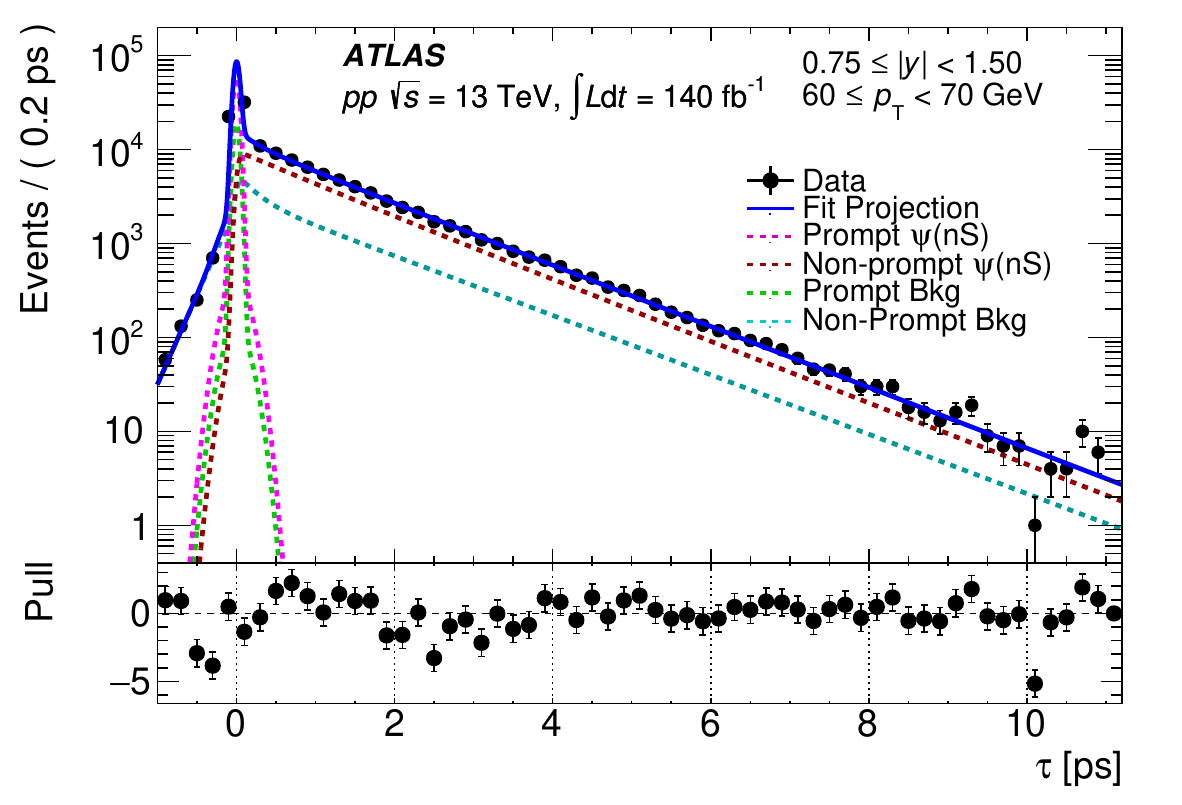}}
\subfigure[]{\includegraphics[width=.43\columnwidth]{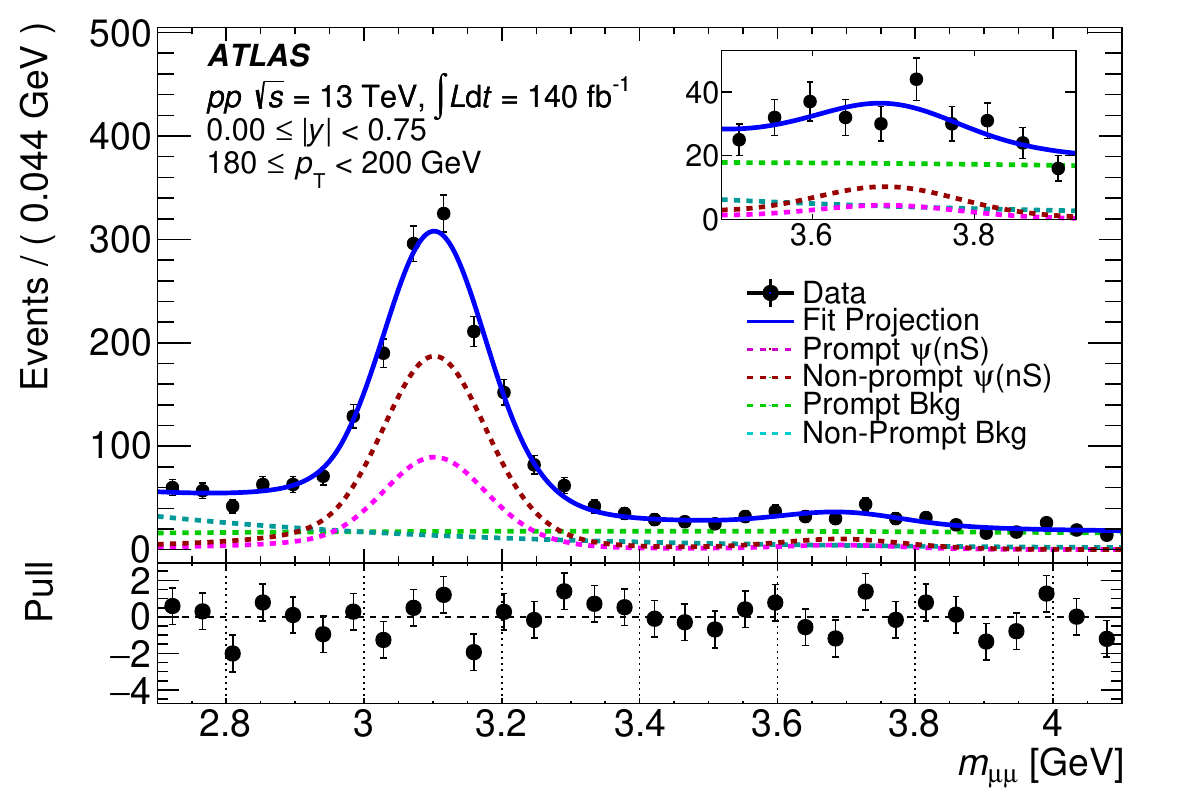}}
\subfigure[]{\includegraphics[width=.43\columnwidth]{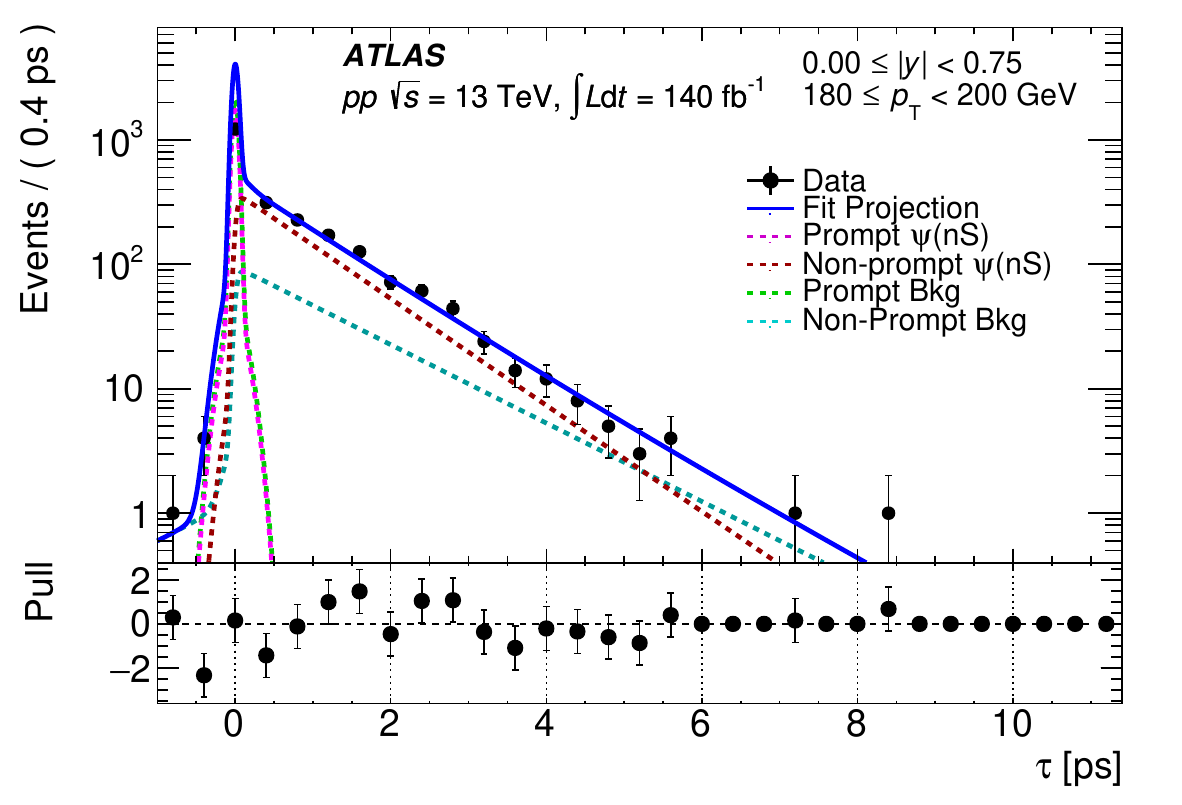}}
\caption{Mass (left) and pseudo-proper decay time (right) projections of the fit result for selected analysis (sub-)bins.
The values of $\chi^2/$d.o.f. for the corresponding 2-dimensional fits are, from top to bottom: 1.09, 0.89, 1.20 and 0.93.
\label{fig:example_fits}}
\end{figure}

Figure~\ref{fig:example_fits} shows the mass and pseudo-proper decay time
projections of the fits in several sample bins, together with the associated pull distributions.
The quality of the fits, assessed by calculating a two-dimensional $\chi^2$ value, is found to be good in all (sub-)bins.

The main parameters determined from the fits are the prompt and non-prompt yields of $\jpsi$ and $\psip$ states. The cross-sections, non-prompt fractions and production ratios were then calculated using Eqs.~(\ref{eq:cross_section})--(\ref{eq:prod_ratio}).
taking into account correlations between fit parameters.
The results for all measured quantities are presented in Section~\ref{sec:results}.

\subsection{Efficiency corrections}
\label{sec:eff_corr}
 
As shown in Eq.~(\ref{eq:cross_section}), the yields obtained from two-dimensional maximum-likelihood fits in each (sub-)bin are subject to acceptance corrections, followed by the corrections for reconstruction and trigger efficiencies obtained from $\jpsi$ and $\psip$ MC simulations, and by the correction scale factors that account for differences between data and MC simulation.
 
MC samples used for efficiency determinations were produced either by the \PYTHIA[8] generator~\cite{Sjostrand:2014zea} with the A14 set of tuned parameters~\cite{ATL-PHYS-PUB-2014-021}, or by a custom particle gun generator producing single $\psi$ states with a given distribution, followed by their decay into a dimuon final state.  The generated events were passed through the full ATLAS detector simulation~\cite{SOFT-2010-01} based on \GEANT~\cite{Agostinelli:2002hh}, and were reconstructed using the same software as the real data.
The reconstruction efficiency $\epsilon_{\mathrm{reco}}$ in each analysis (sub-)bin is defined
as the ratio
\begin{equation*}
\epsilon_{\mathrm{reco}} = \frac{N_\mathrm{reco}}{N_\mathrm{true}},
\end{equation*}
where $N_\mathrm{reco} $ is the reconstructed yield within the bin boundaries defined in terms of reconstructed variables, with fiducial cuts applied to reconstructed variables, while $N_\mathrm{true}$
is the true (generated) yield within the bin boundaries defined in terms of true variable values, with fiducial cuts applied to the true variable values.
This definition takes into account detector resolution smearing of the kinematic variables used to define the fiducial cuts and bin boundaries, and hence includes bin migration effects between neighbouring bins.
 
The trigger efficiency $\epsilon_{\mathrm{trig}}$, also obtained using MC simulations, is defined fully in terms of reconstructed variables as
\begin{equation*}
\epsilon_{\mathrm{trig}} = \frac{N_\mathrm{trig}}{N_\mathrm{reco}},
\end{equation*}
where $N_\mathrm{trig} $ is the number of triggered events among the reconstructed events.
Since this measurement used two different triggers, the trigger efficiency was calculated accordingly.
Further correction scale factors, $\epsilon_{\mathrm{recoSF}}$ and $\epsilon_{\mathrm{trigSF}}$, are applied to account for any differences between data and MC events at reconstruction level and trigger level
(see Eq.~(\ref{eq:cross_section})). These are evaluated
in dedicated tag-and-probe studies with auxiliary triggers, using a mixture of $\jpsi \to \mu^+\mu^-$ and $Z \to \mu^+\mu^-$ decays
(see Ref.~\cite{PERF-2015-10} for details).

% End of text imported from the .//AnalysisStrategy.tex input file

\section{Systematic uncertainties}
\label{sec:systematics}

% The next lines are included from the .//SystematicStudies.tex input file
Systematic effects from a variety of sources were studied, and appropriate corrections and uncertainties were assigned to all measured quantities. The systematic uncertainties can be broadly grouped into those related to reconstruction, trigger, and acceptance corrections, and those related to the fit model.
 
In order to assess the systematic uncertainties related to the fit model, the fit model was varied in several ways. As mentioned in Section~\ref{ssec:fit}, in the nominal fits some of the parameters describing the lineshapes of the signal peaks in the mass and lifetime domains were fixed to the values obtained from signal MC studies. These include the values of parameters $\alpha$ and $n$ of the Crystal Ball function, the constant $\beta$ linking the positions of the $\jpsi$ and $\psip$ mass peaks, and the factor $b$ relating the slopes of the two exponential functions describing the distribution of $\jpsi$ decay times, $\tau$.
These parameters were allowed to float,
one at a time, and the fits were repeated.
Some variations covered changes in the decay time resolution parameterisation, with widths of the three Gaussian functions changed independently. In other variations, alternative parameterisations were chosen for the mass dependence of the background terms, and the fits were repeated. In one of the variations the correlation parameter $\rho$ from Eq.~\ref{eq:model2} was set to 0.
Finally, in the highest $\pt$  bins, $\pt>140$~\GeV, the fixed fraction 0.07, relating the $\psip$ and $\jpsi$ yields, was varied by 0.01 to cover the observed range at lower $\pt$ values, and the fits were run again.
After each rerun, changes in the measured yields
were recorded. The outcome of this process, in each analysis bin, was a number of measurements of each yield, scattered around the result of the nominal fit for that yield.
It was assumed that the probability distribution
of these variations around the nominal value was uniform between the smallest and largest measured values, and, therefore, the corresponding systematic uncertainty was evaluated as the standard deviation of this uniform distribution.
 
Acceptance-related systematic uncertainties are governed by the size of the samples used to generate the corresponding acceptance maps. The chosen size ensured that these uncertainties are small relative to the total systematic uncertainties.
Changes in acceptance due to different spin-alignment hypotheses were treated separately (see Appendix).
 
Systematic uncertainties due to trigger and reconstruction efficiency corrections
include the uncertainties on the scale factors~\cite{PERF-2015-10} as well as those due the sizes and shapes of the MC samples. These were added in quadrature.
 
The total systematic uncertainty in each bin is calculated by summing in quadrature the uncertainties from the above-mentioned  sources.
The total uncertainty is calculated as the sum in quadrature of the total systematic and statistical uncertainties.
 
An additional systematic uncertainty comes from the determination of the integrated luminosity.
In the $\pt < 60~\GeV$ region, where only 2015 data contributes to this measurement, the integrated luminosity has a 1.13\% uncertainty~\cite{DAPR-2021-01}, while for $\pt \ge 60~\GeV$ the uncertainty in the combined 2015--2018 integrated luminosity is 0.83\%, as obtained using the LUCID-2 detector~\cite{LUCID2}.

\begin{figure}[H]
\subfigure[]{\includegraphics[width=.49\columnwidth]{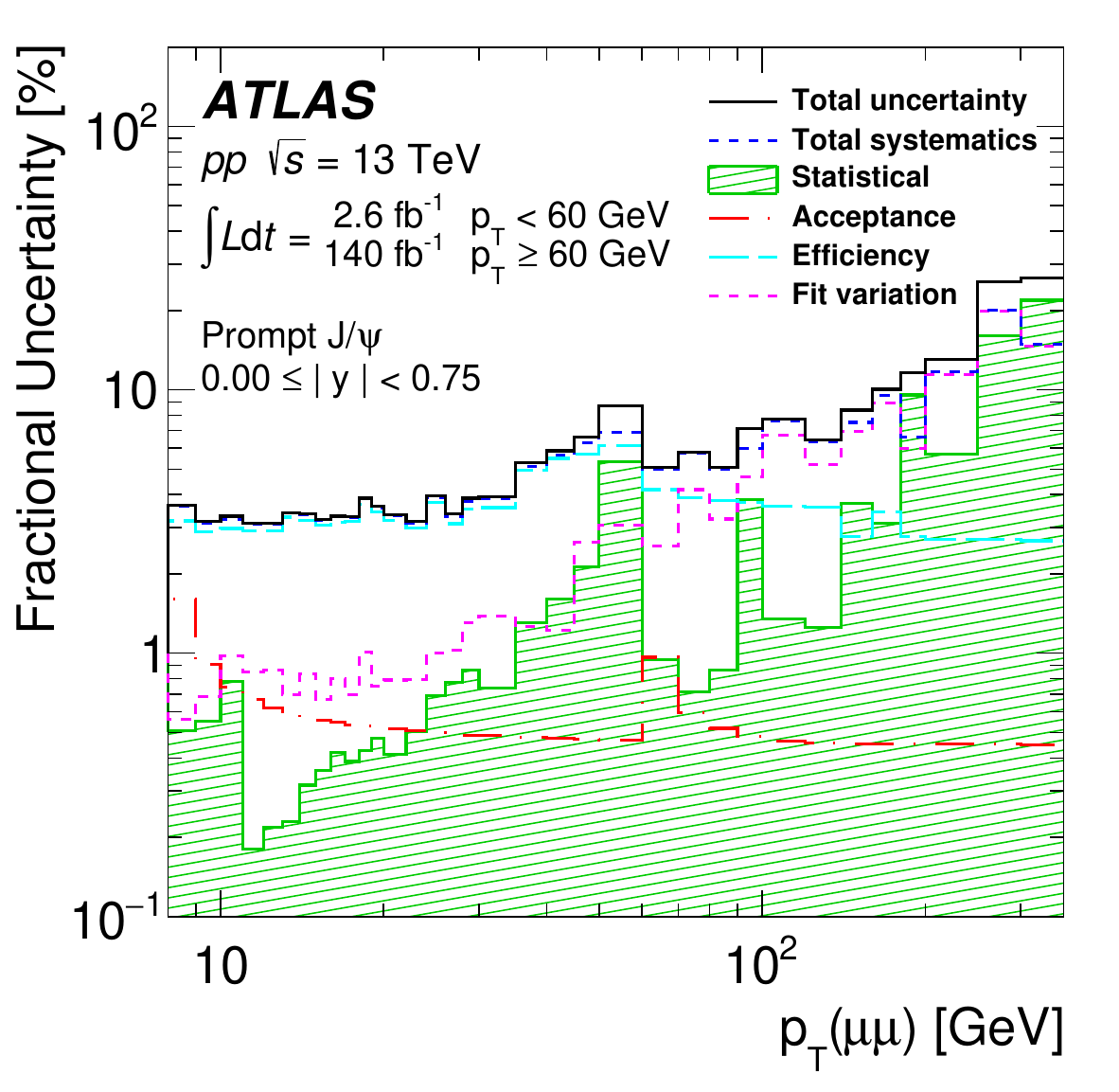}}
\subfigure[]{\includegraphics[width=.49\columnwidth]{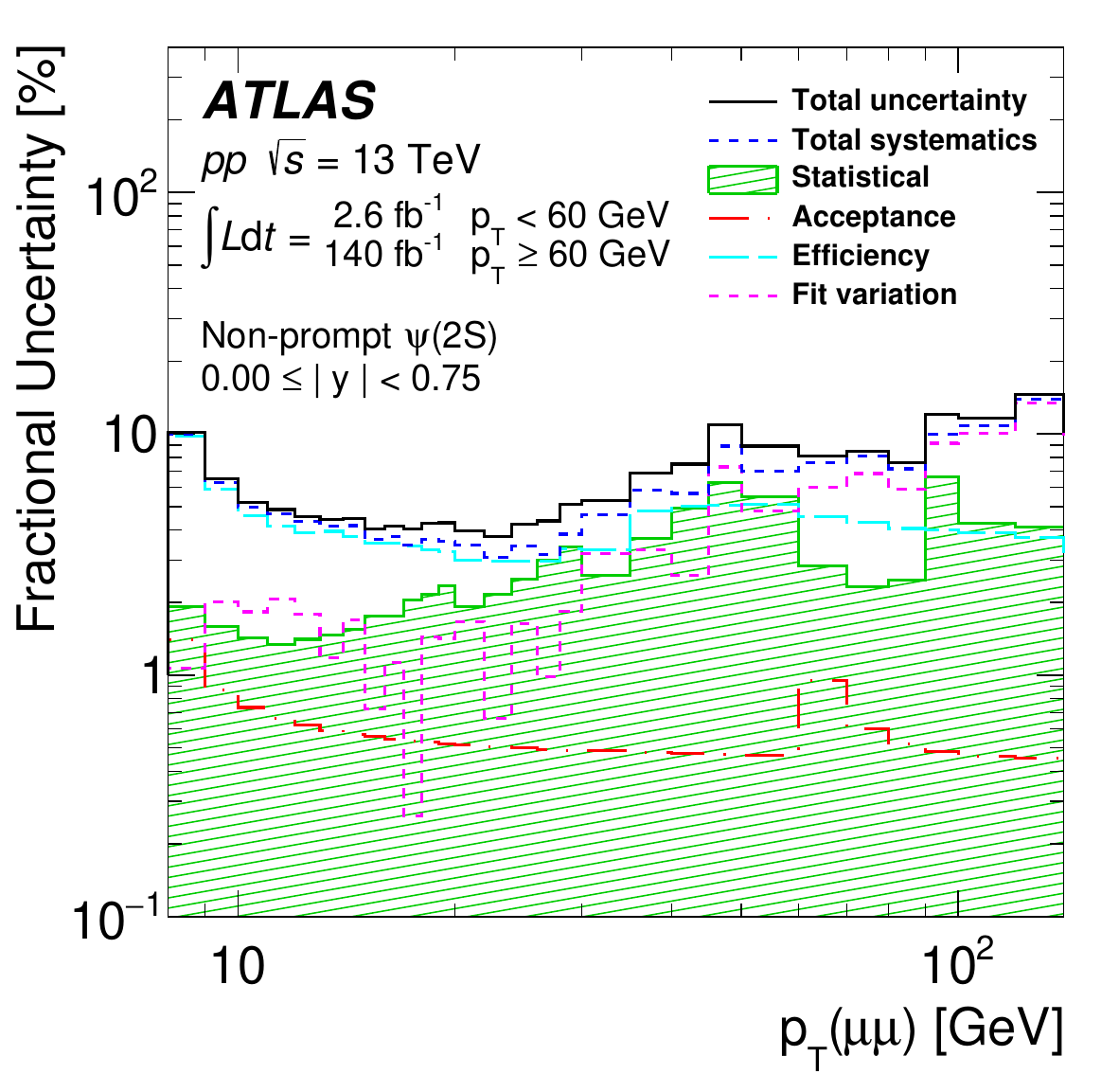}}
\subfigure[]{\includegraphics[width=.49\columnwidth]{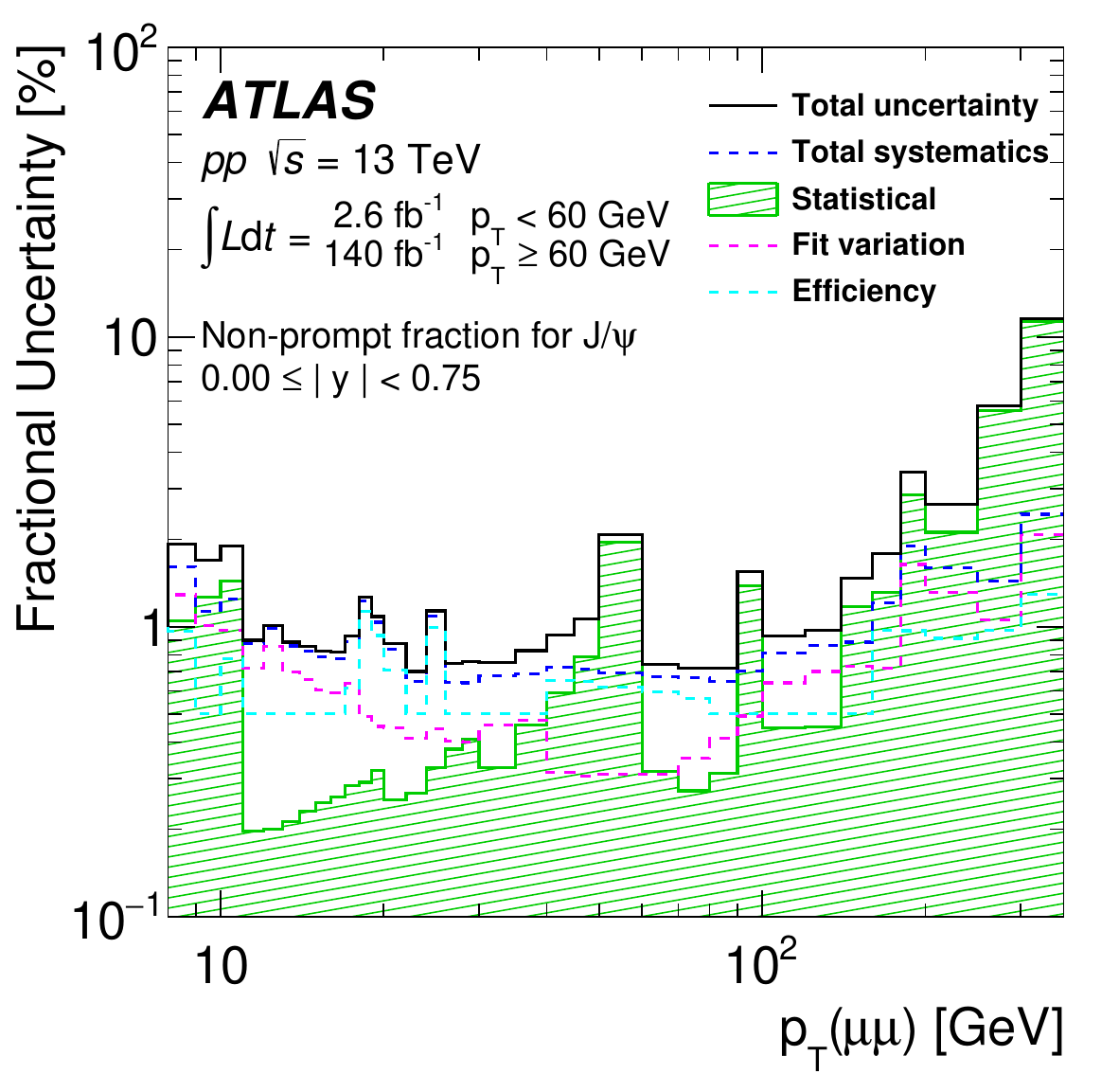}}
\subfigure[]{\includegraphics[width=.49\columnwidth]{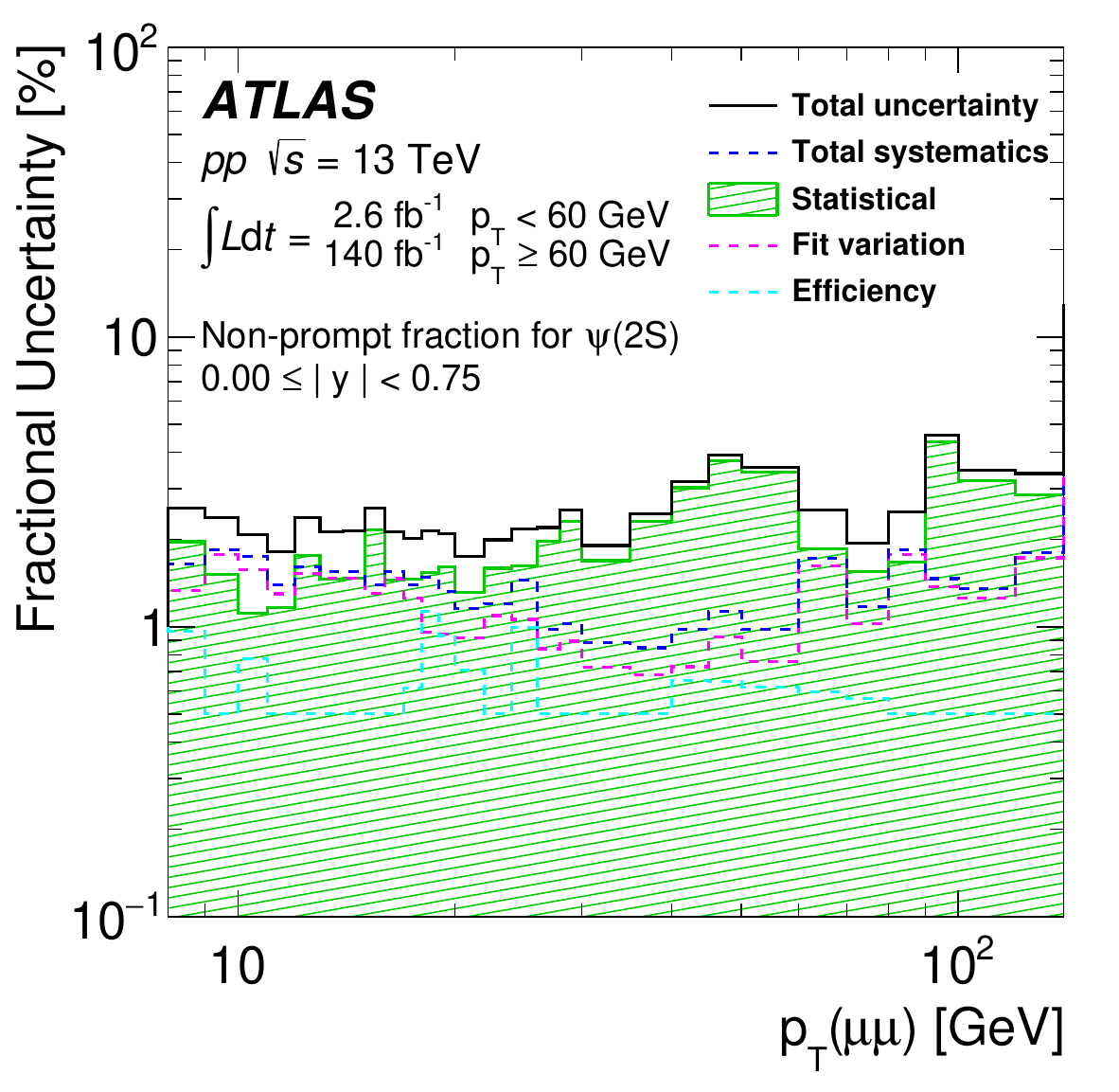}}
\caption{Total, statistical, and systematic uncertainties (in \%) as functions of $\pt$ for the differential (a) prompt $\jpsi$ and (b) non-prompt $\psip$  cross-sections, and for the non-prompt fractions of (c) $\jpsi$ and (d) $\psip$, in the rapidity slice $0.00 \leq |y|<0.75$. The main components of the systematic uncertainties are also shown.}
\label{fig:systs}
\end{figure}

As an illustration, in Figure~\ref{fig:systs} the total, statistical, and systematic uncertainties, together with the main individual contributions to the systematic uncertainty, are shown for the central rapidity slice for the differential cross-sections of prompt $\jpsi$ and non-prompt $\psip$ mesons, as well as for the non-prompt fractions of $\jpsi$ and $\psip$ mesons, as functions of $\pt$. For the cross-sections, apart from a few high $\pt$ bins, the uncertainties are largely systematic, while for the non-prompt fractions and $\psip$-to-$\jpsi$ production ratios, statistical errors dominate in many bins because the systematic uncertainties partially cancel out.
Error levels are similar between prompt and non-prompt cross sections. Comparable results are obtained in other rapidity slices.

% End of text imported from the .//SystematicStudies.tex input file

\section{Results}
\label{sec:results}

\begin{figure}[htb]
\subfigure[]{\includegraphics[width=.5\columnwidth]{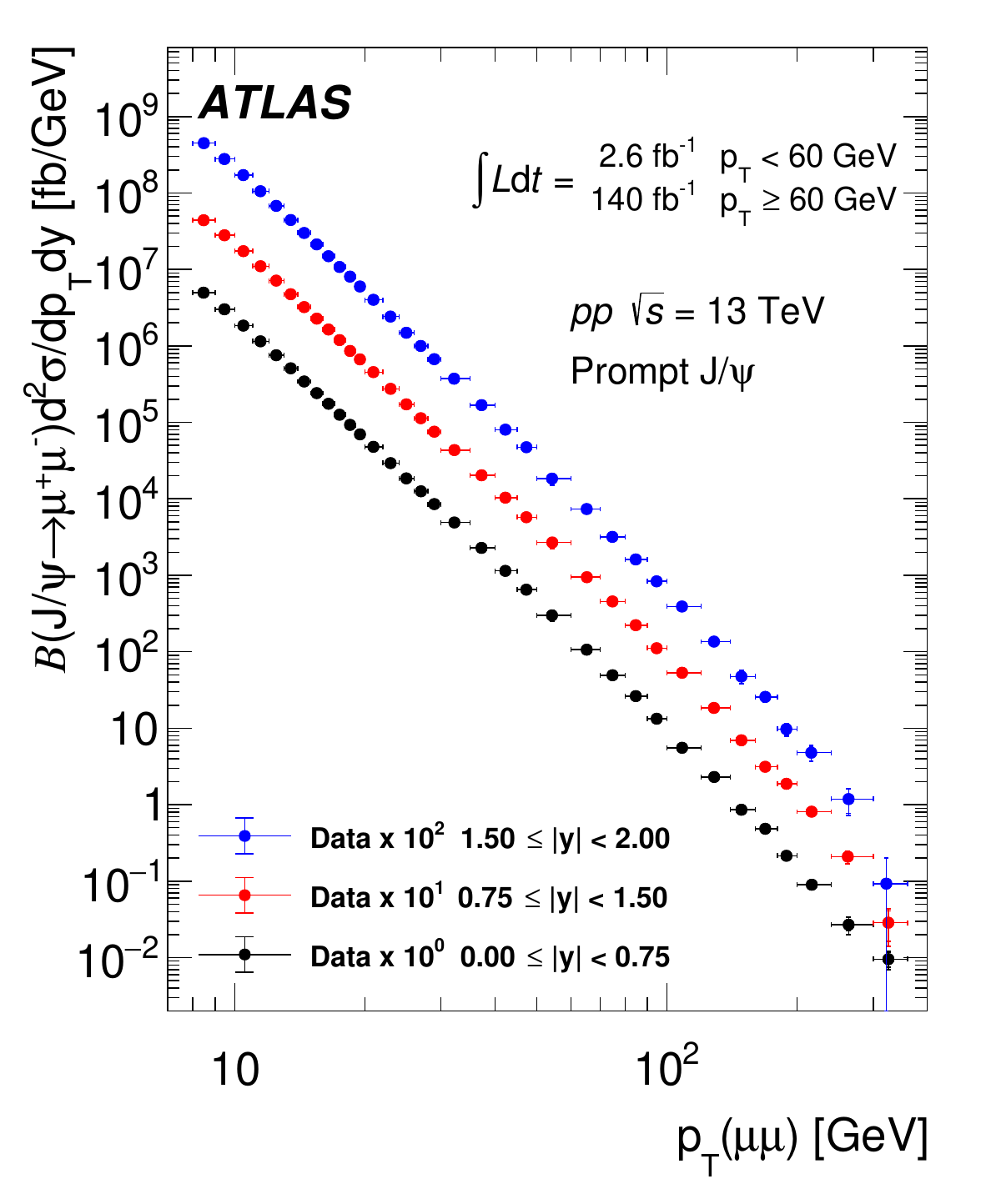}}
\subfigure[]{\includegraphics[width=.5\columnwidth]{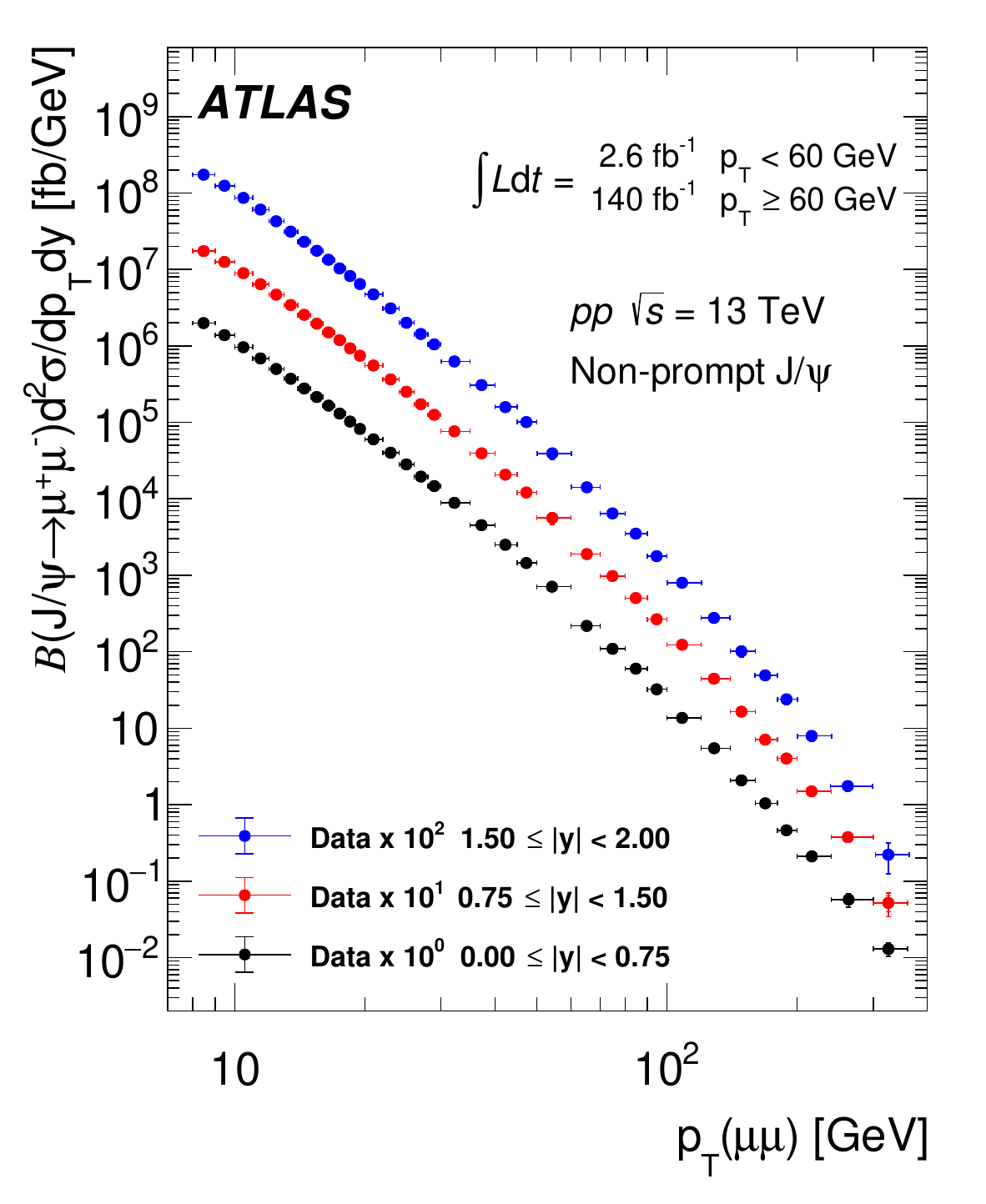}}
\caption{Differential cross-sections for (a) prompt and (b) non-prompt production of $\jpsi$ mesons.
For visual clarity, a scaling factor of $1$, $10$, or $100$ is applied to the rapidity slices $0.00 \leq |y|<0.75$, $0.75\le |y|<1.5$, and $1.5\le |y|<2.0$, respectively.
For each data point, the horizontal bar spans the $\pt$ range covered by that bin, with the horizontal position of each point representing the mean $\pt$ in that bin. The vertical uncertainty range (obscured by the marker for some values) combines both the statistical (the inner bar) and total uncertainty. Uncertainties related to spin alignment or integrated luminosity are not included.
Data up to $60~\si{\gev}$ were taken with a dimuon trigger with integrated luminosity $2.6\,\si{fb}^{-1}$;
data above $60~\si{\gev}$ were taken with a single-muon trigger with integrated luminosity $140\,\si{fb}^{-1}$.
}
\label{fig:jpsi_cross}
\end{figure}

\begin{figure}[hbt]
\subfigure[]{\includegraphics[width=.5\columnwidth]{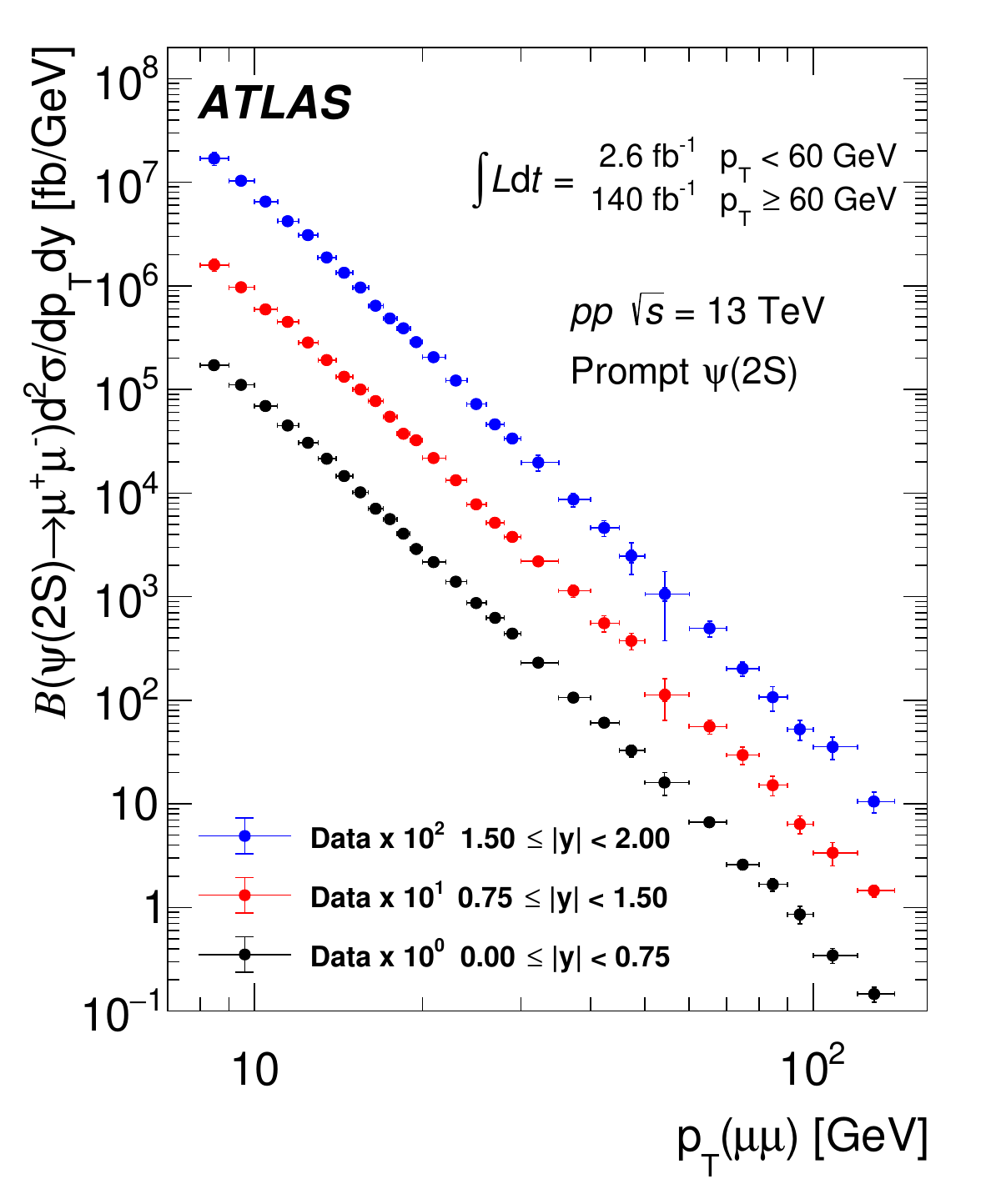}}
\subfigure[]{\includegraphics[width=.5\columnwidth]{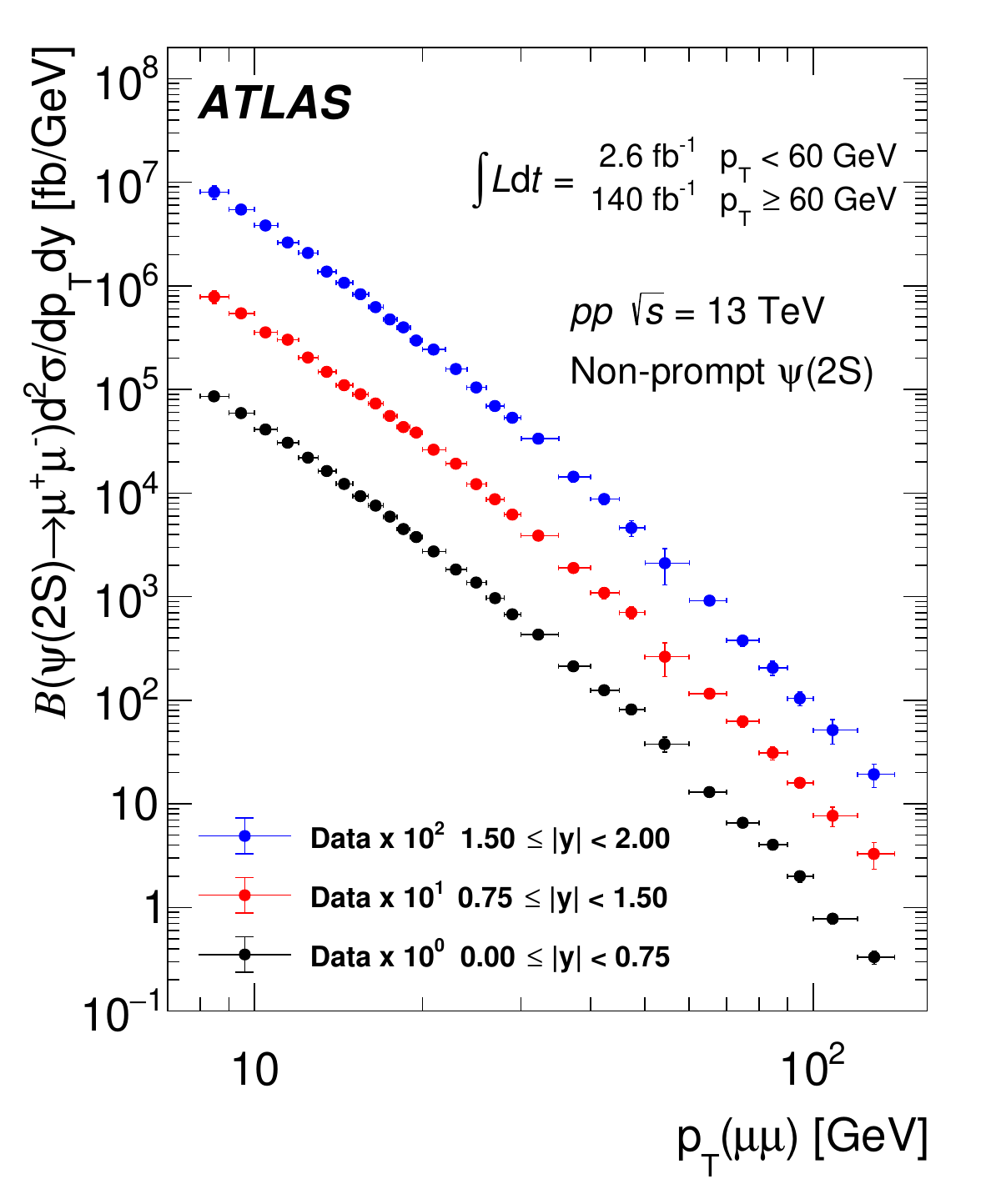}}
\caption{Differential cross-sections for (a) prompt and (b) non-prompt production of
$\psip$ mesons.
For visual clarity, a scaling factor of $1$, $10$, or $100$ is applied to the rapidity slices $0.00 \leq |y|<0.75$, $0.75\le|y|<1.5$, and $1.5\le|y|<2.0$, respectively.
For each data point, the horizontal bar spans the $\pt$ range covered by that bin, with the horizontal position of each point representing the mean $\pt$ in that bin. The vertical uncertainty range (obscured by the marker for some values) combines both the statistical (the inner bar) and total uncertainty.
Uncertainties related to spin alignment or integrated luminosity are not included.
Data up to $60~\si{\gev}$ were taken with a dimuon trigger with integrated luminosity $2.6\,\si{fb}^{-1}$;
data above $60~\si{\gev}$ were taken with a single-muon trigger with integrated luminosity $140\,\si{fb}^{-1}$.
}
\label{fig:psip_cross}
\end{figure}

\begin{figure}[htb]
\subfigure[]{\includegraphics[width=.5\columnwidth]{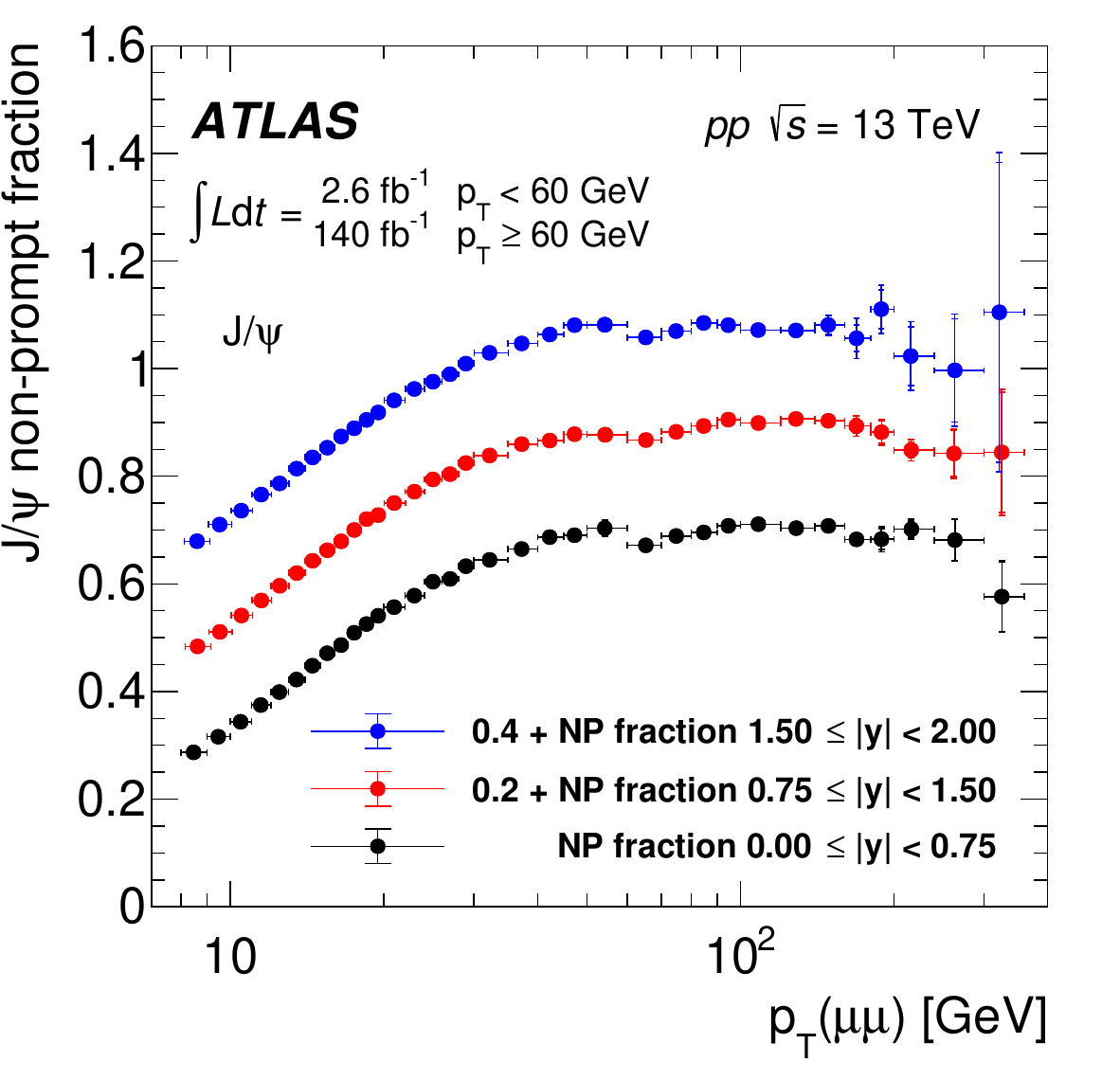}}
\subfigure[]{\includegraphics[width=.5\columnwidth]{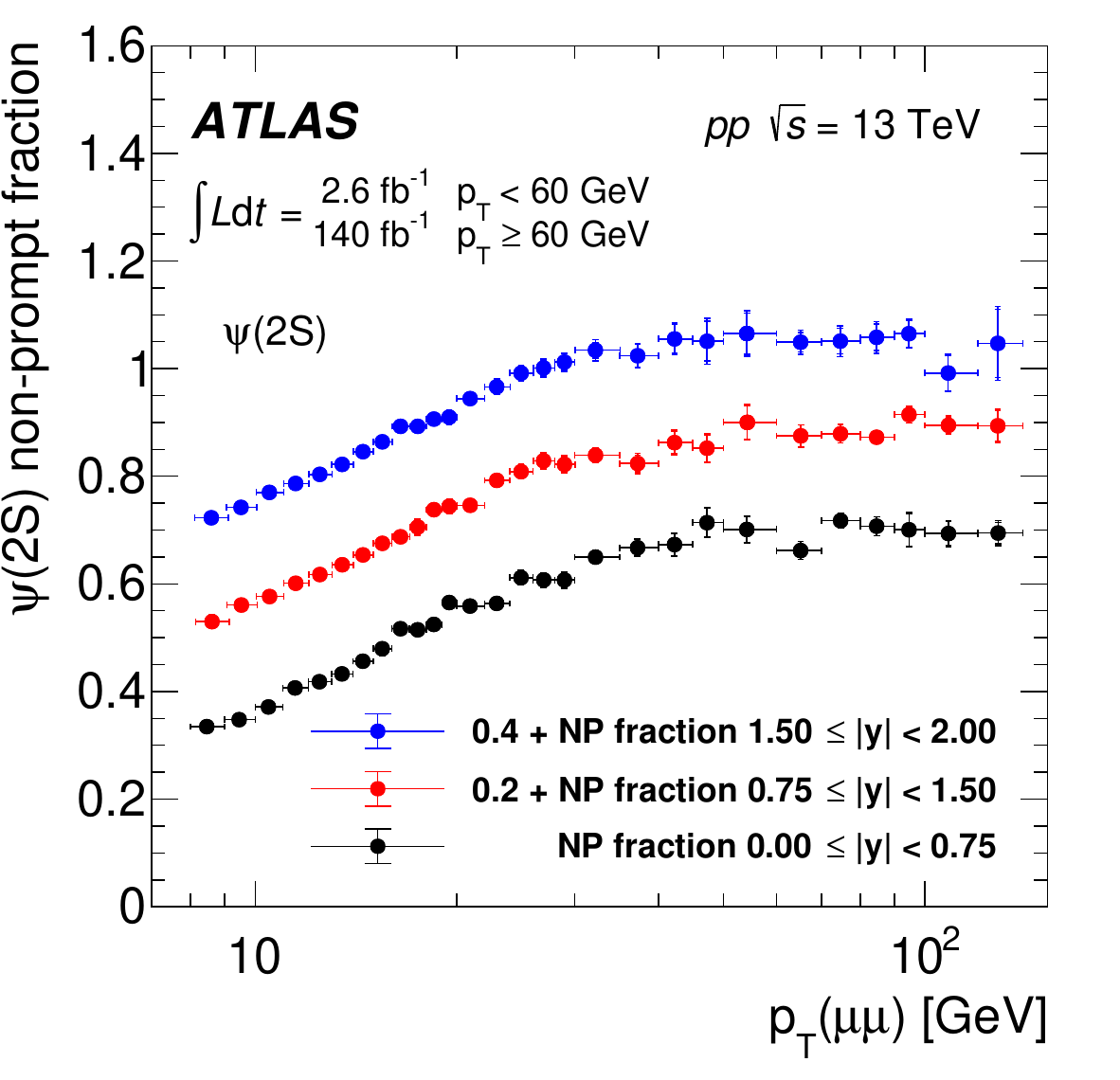}}
\caption{Non-prompt production fraction of (a) $\jpsi$ and (b) $\psip$ mesons.
For visual clarity, a vertical shift of $0$, $0.2$, or $0.4$ is applied to the rapidity slices $0.00 \leq |y|<0.75$, $0.75\le|y|<1.5$, and $1.5\le|y|<2.0$, respectively.
For each data point, the horizontal bar spans the $\pt$ range covered by that bin, with the horizontal position of each point representing the mean $\pt$ in that bin. The vertical uncertainty range (obscured by the marker for some values) combines both the statistical (the inner bar) and total uncertainty. Uncertainties related to spin alignment or integrated luminosity are not included.
Data up to $60~\si{\gev}$ were taken with a dimuon trigger with integrated luminosity $2.6\,\si{fb}^{-1}$;
data above $60~\si{\gev}$ were taken with a single-muon trigger with integrated luminosity $140\,\si{fb}^{-1}$.
}
\label{fig:npfrac}
\end{figure}

\begin{figure}[htb]
\subfigure[]{\includegraphics[width=.5\columnwidth]{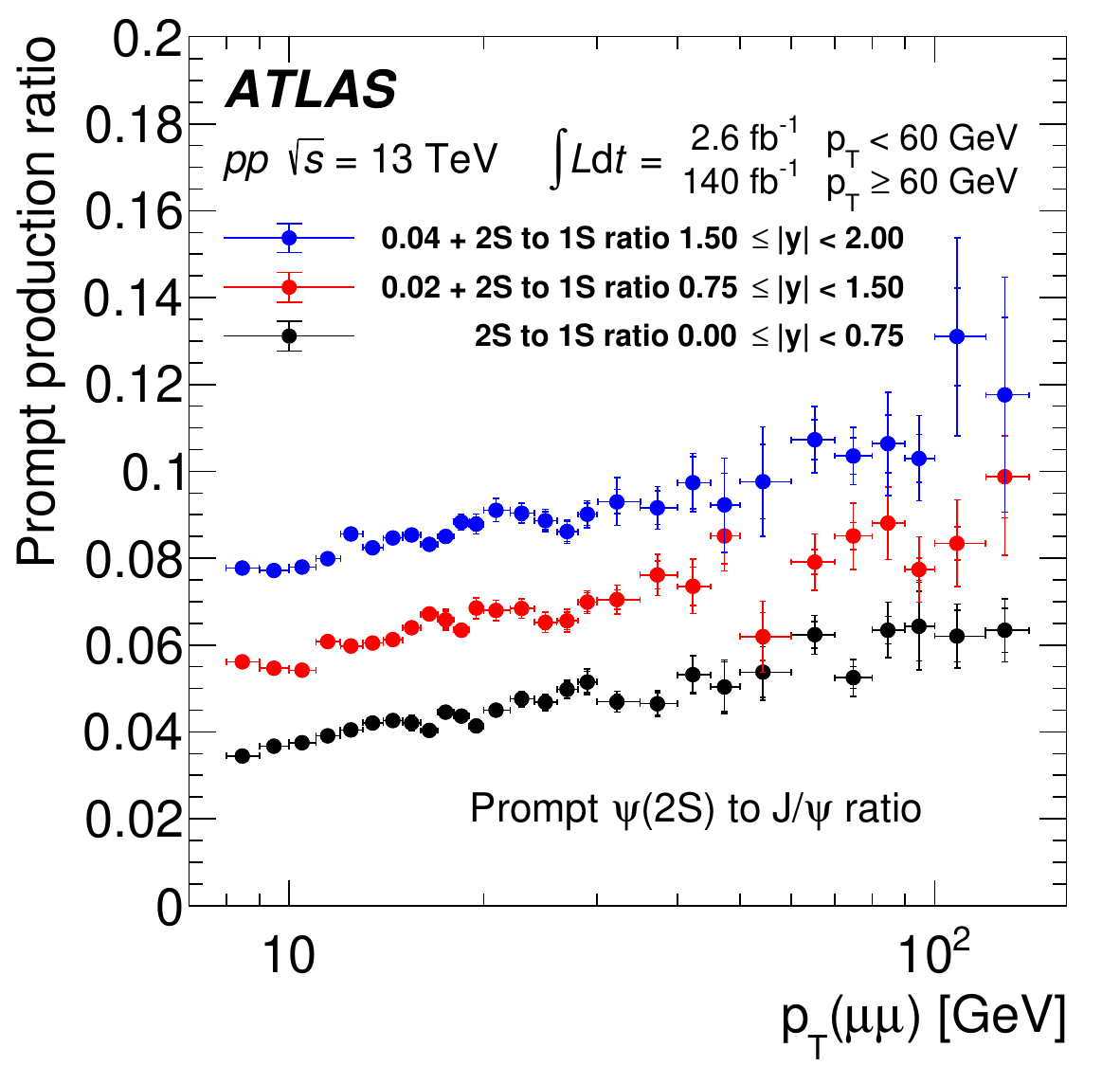}}
\subfigure[]{\includegraphics[width=.5\columnwidth]{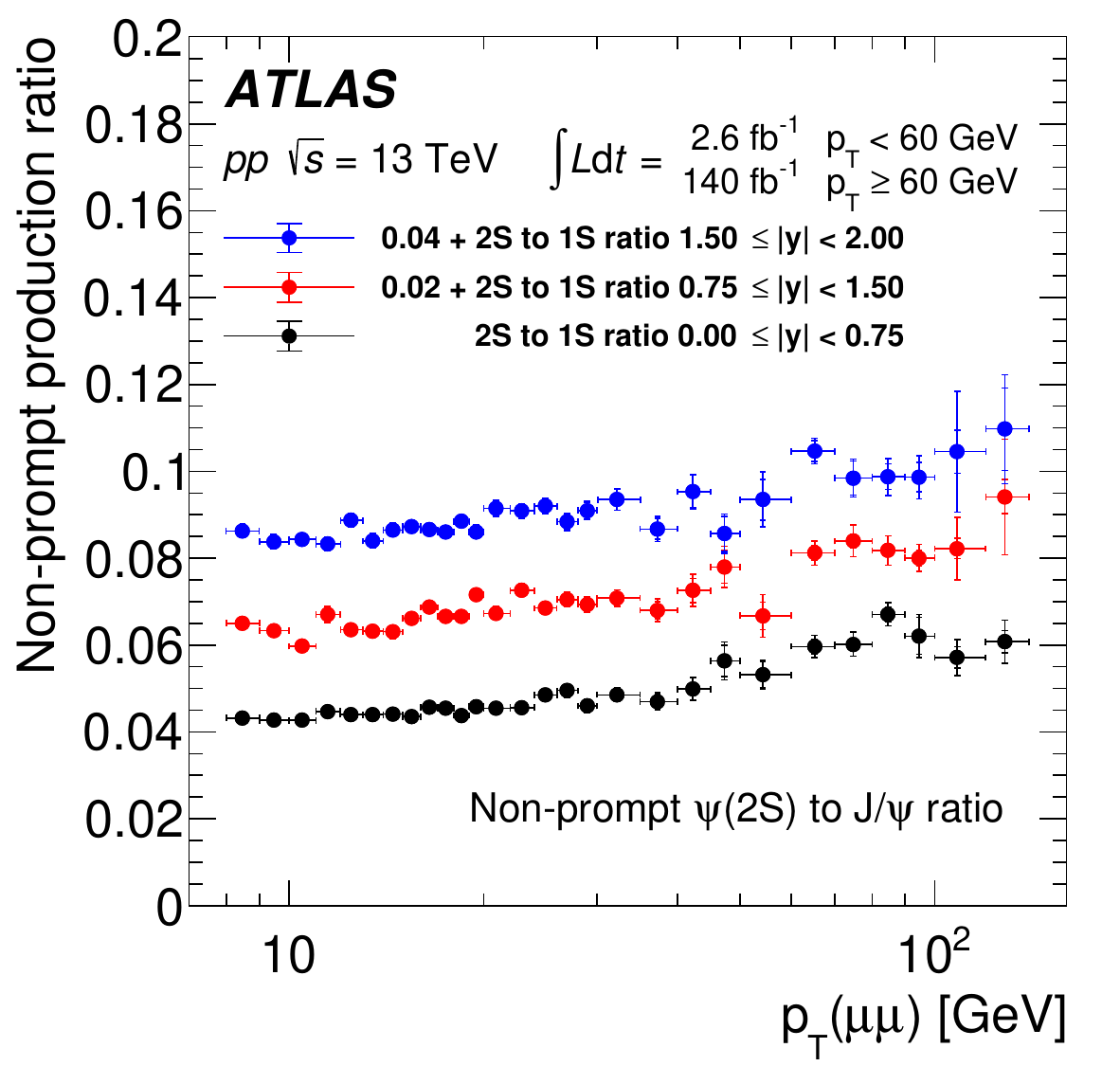}}
\caption{The $\psip$-to-$\jpsi$ production ratio for the (a) prompt  and
(b) non-prompt  production mechanisms.
For visual clarity, a vertical shift of $0$, $0.02$, or $0.04$ is applied to the rapidity slices $0.00 \leq |y|<0.75$, $0.75\le|y|<1.5$, and $1.5\le|y|<2.0$, respectively.
For each data point, the horizontal bar spans the $\pt$ range covered by that bin, with the horizontal position of each point representing the mean $\pt$ in that bin. The vertical uncertainty range (hidden by the marker for some values) combines both the statistical (the inner bar) and total uncertainty. Uncertainties related to spin alignment or integrated luminosity are not included.
Data up to $60~\si{\gev}$ were taken with a dimuon trigger with integrated luminosity $2.6\,\si{fb}^{-1}$;
data above $60~\si{\gev}$ were taken with a single-muon trigger with integrated luminosity $140\,\si{fb}^{-1}$.
}
\label{fig:ratio}
\end{figure}

% The next lines are included from the .//Results.tex input file
The measured double-differential cross-sections for prompt and non-prompt $\jpsi$ production in the nominal isotropic spin-alignment scenario are presented in
Figures~\ref{fig:jpsi_cross}(a) and~\ref{fig:jpsi_cross}(b), respectively.
The same quantities for $\psip$ are shown in
Figures~\ref{fig:psip_cross}(a) and~\ref{fig:psip_cross}(b).
The non-prompt production fractions for $\jpsi$ and $\psip$ are presented in
Figures~\ref{fig:npfrac}(a) and~\ref{fig:npfrac}(b).
Finally, the $\psip$-to-$\jpsi$ production ratios are presented in
Figures~\ref{fig:ratio}(a) and~\ref{fig:ratio}(b) for the prompt and non-prompt production mechanisms, respectively.
 
While the non-prompt fractions shown in Figure~\ref{fig:npfrac} increase steadily with $\pt$ up to about 100~\GeV,  they are almost constant for both $\jpsi$ and $\psip$ in the high $\pt$ range, which suggests similar $\pt$-dependences for the prompt and non-prompt differential cross-sections at very high transverse momenta.

\subsection{Acceptance and spin alignment corrections}

The transition between the low-$\pt$ dimuon trigger and the high-$\pt$ single-muon trigger at $\pt=60$~\GeV\ presents a particular challenge because of the sharp change in event kinematics. The corresponding changes in the acceptance and efficiency correction factors are significant and could lead to discontinuities in the measured distributions.
 
\begin{figure}[htb]
\subfigure[]{\includegraphics[width=.5\columnwidth]{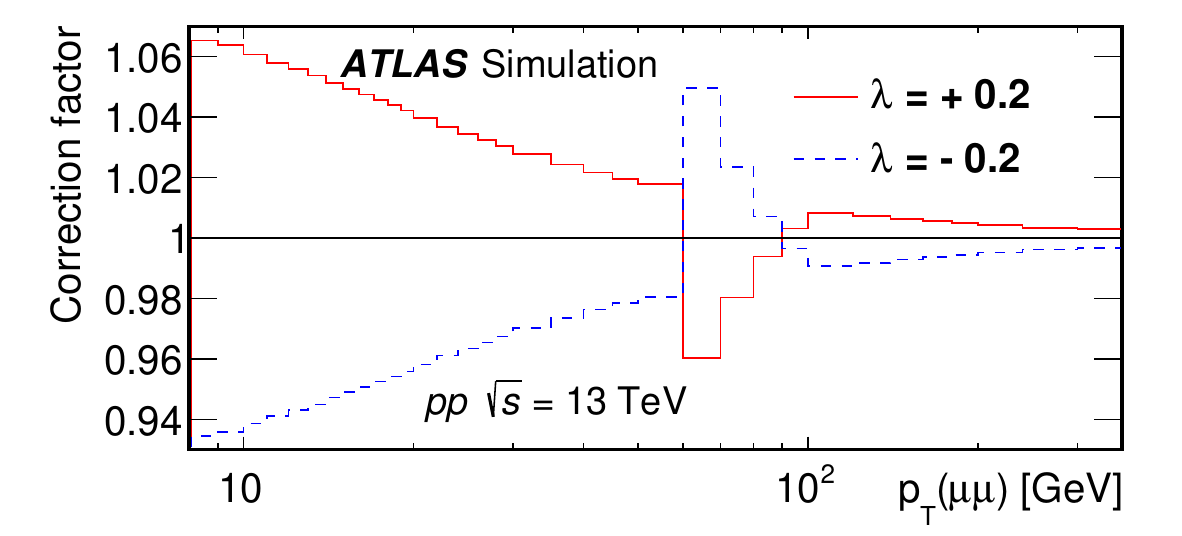}}
\subfigure[]{\includegraphics[width=.5\columnwidth]{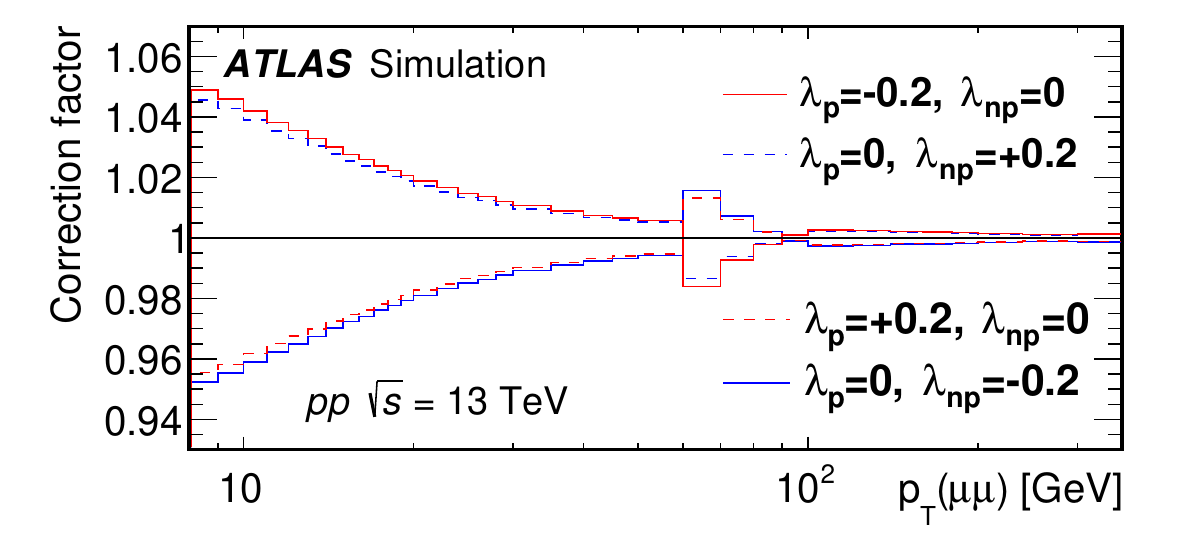}}
\caption{Spin-alignment hypothesis correction factors for the $\jpsi$ (a) differential cross-section and (b) non-prompt production fraction, for a number of spin-alignment scenarios. The correction factors are approximately the same for $\jpsi$ and $\psip$, for the prompt and non-prompt production mechanisms, and also for the three rapidity regions. The discontinuities at $\pt=60$~\GeV\ are due to the transition from a low-$\pt$ dimuon trigger to a high-$\pt$ single-muon trigger, and the corresponding change in event acceptance.}
\label{fig:Spin_corr}
\end{figure}

Since the spin alignment of $\psi$ states may be different for the prompt and non-prompt production mechanisms, additional correction factors may be needed for all measured distributions.
In order to quantitatively assess the possible impact of $\psi$ spin alignment, correction factors are calculated for a variety of scenarios. It was found that the dependence on the  polar angle $\theta$ in the helicity frame of the $\psi$ state causes the largest variation, so
the angular dependence of $\psi\to \mu^+\mu^-$ decays is assumed to be $\propto(1+\lambda_{\theta}\cos^2\theta)$.
The correction factors are shown in
Figures~\ref{fig:Spin_corr}(a) and~\ref{fig:Spin_corr}(b) for the differential cross-sections and non-prompt fractions respectively,
where the values $\lambda_{\theta} = \pm 0.20$ are chosen to reflect the approximate level of experimental
knowledge~\cite{Chatrchyan:2011kc,PhysRevLett.85.2886,ALICE:2011gej}
and theoretical understanding~\cite{Ma:2018qvc, Feng:2018ukp,Faccioli:2022alj} of this parameter.
The correction factors are shown for prompt $\jpsi$ in the central rapidity range, but were found to be essentially the same for
$\jpsi$ and $\psip$, for the prompt and non-prompt production mechanisms, and also for the three rapidity regions.
 
The potential bias due to the spin-alignment assumption is especially noticeable at the $\pT = 60$~\GeV\ transition, and indeed a step can be seen at this point in the $\jpsi$ non-prompt fraction in Figure~\ref{fig:npfrac}(a), reflecting a possible issue in the spin-alignment modelling.
Correction factors for some other values of $\lambda_{\theta}$ are presented in the Appendix. These can be used for further studies, if more precise spin-alignment data and/or improved modelling become available in the future.
 
\subsection{Theory comparison: prompt production}
 
Model calculations of prompt production of charmonium are usually based on perturbative QCD for the production of the $c\bar c$ pair, and differ in the mechanism of hadronisation and formation of the bound state with specific quantum numbers.

\begin{figure}[h]
\subfigure[]{\includegraphics[width=.49\columnwidth]{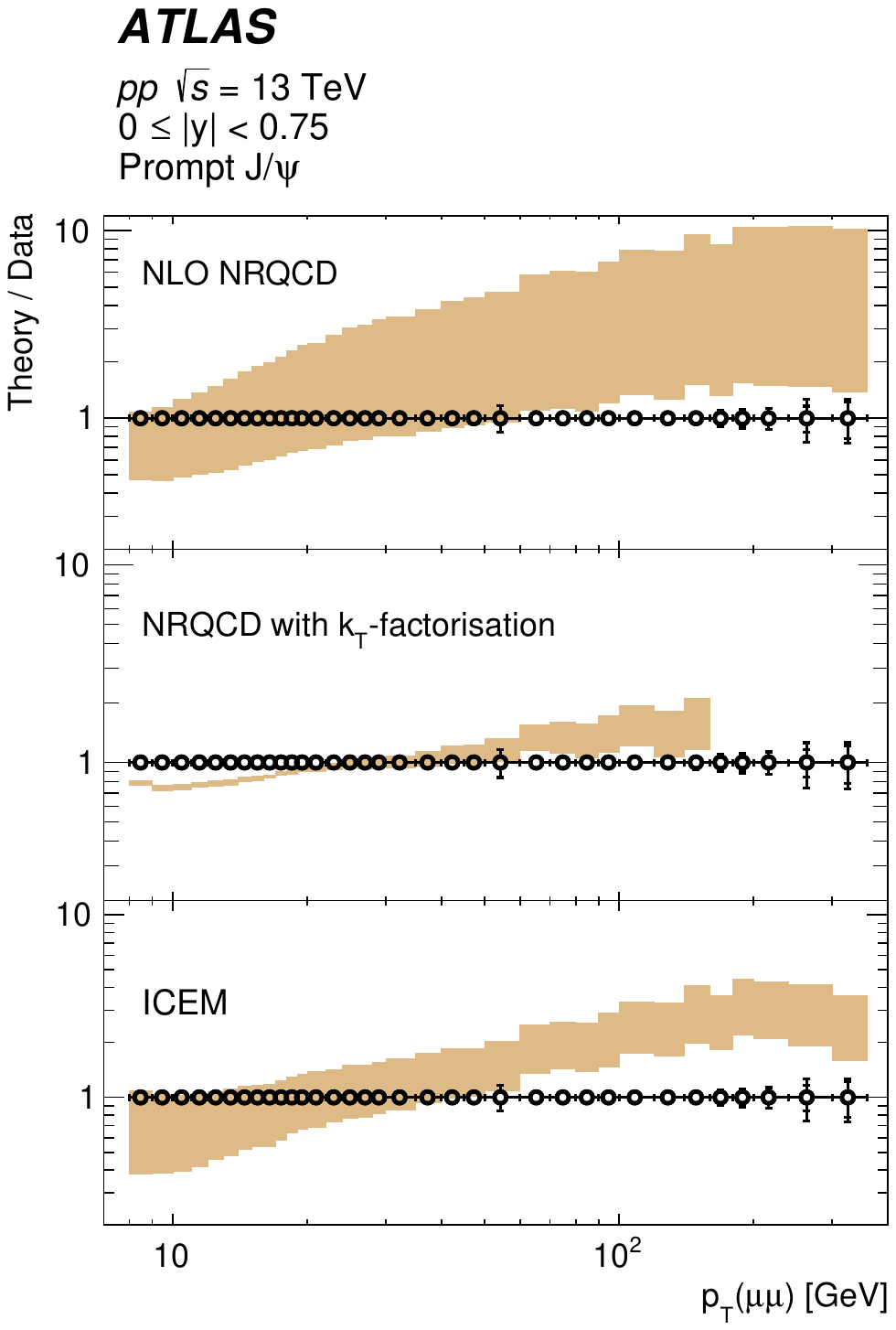}}
\subfigure[]{\includegraphics[width=.49\columnwidth]{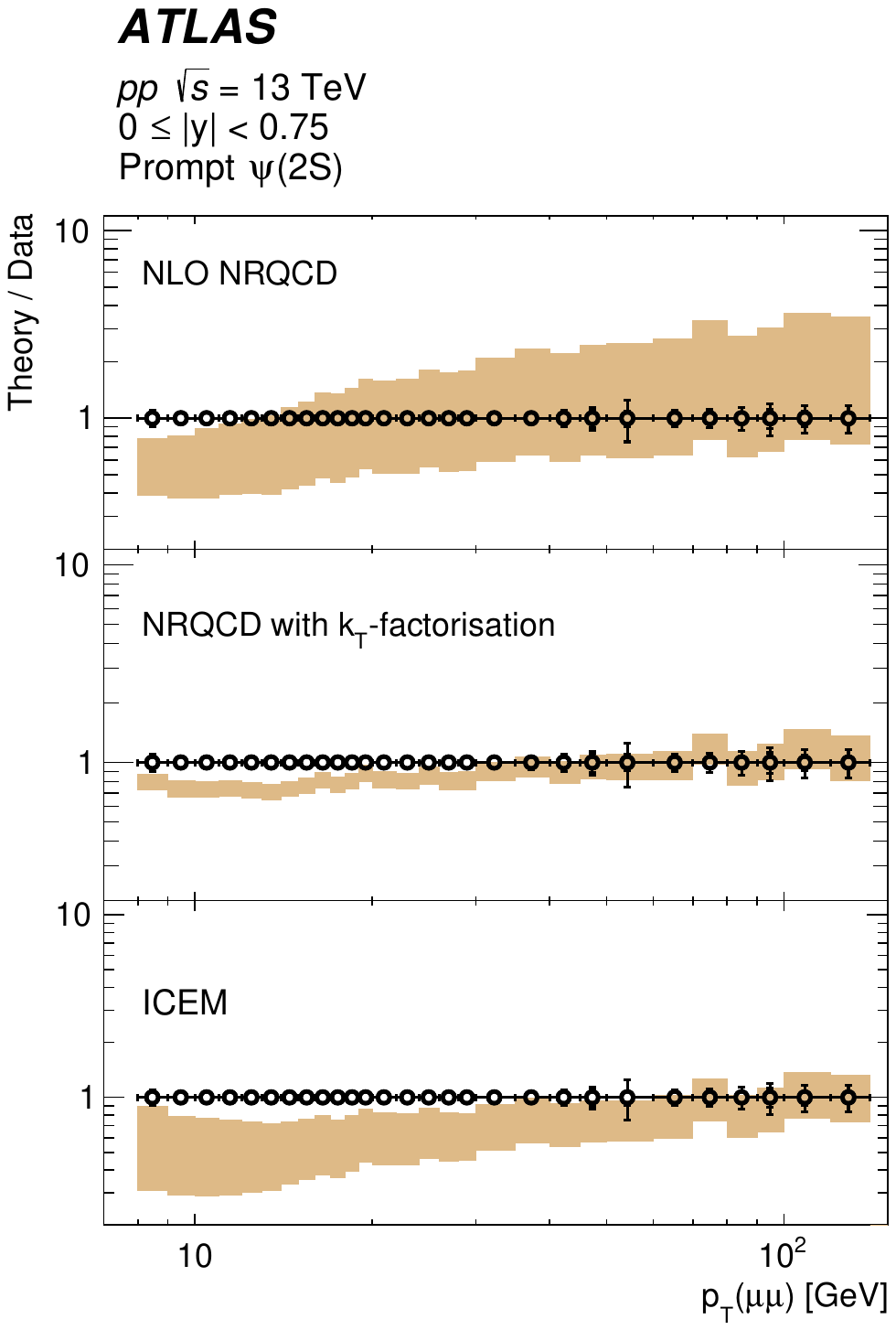}}
\caption{Ratios of various theoretical predictions (described in the text) to the data points from this measurement, for the prompt production of (a) $\jpsi$ and (b)  $\psip$ in the central rapidity region. In each $\pt$ bin, the shaded area represents the ratio of the theoretical prediction to the measured value, with the vertical spread showing the uncertainties of the respective model.
Error bars on the black circles show fractional uncertainties of this measurement.
}
\label{fig:Pred_Jpsi_P}
\end{figure}

The predictions of a model using the non-relativistic QCD approach to charmonium production cross-sections at next-to-leading order (NLO NRQCD)~\cite{PRL1062011022003}, using predetermined LDMEs \cite{PRD842011051501,Butenschoen:2022orc}, are shown in comparison with our measurements of the $\jpsi$ and $\psip$ production cross-sections in the top panels of Figures~\ref{fig:Pred_Jpsi_P}(a) and~\ref{fig:Pred_Jpsi_P}(b) respectively.
The predictions of the model largely overlap with the data points within the theoretical uncertainties, which include variations of the renormalisation, factorisation and NRQCD scales.  However, the predictions seem to overestimate the cross-sections at high $\pt$.

One generalisation of the NRQCD approach is a model which aims to improve the description by taking into account the transverse degrees of freedom of the initial gluons in the colliding
protons ($k_T$-factorisation model)~\cite{BaranovEur.Phys.J.C754552015,Phys.Rev.D960340192017}.
The predictions of this model, obtained with
the PEGASUS event generator~\cite{Eur.Phys.J.C803302020} using the LDMEs determined in Ref.~\cite{Phys.Rev.D1001140212019}, are compared
with our measured results as shown in the middle panels of Figures~\ref{fig:Pred_Jpsi_P}(a) and~\ref{fig:Pred_Jpsi_P}(b) for $\jpsi$ and $\psip$ respectively, where the theoretical uncertainties only account for variations of the renormalisation scale.
The predictions of the model tend to underestimate the cross-sections at low $\pt$, while in the high $\pt$ region
the range of comparison is limited by the availability of the transverse-momentum-dependent parton distribution function (PDF) of the gluon~\cite{Jung:2004gs}.
 
A different approach is used by the Improved Colour Evaporation Model (ICEM)~\cite{PhysRevD.104.094026}, which assigns a fixed fraction of the $c\bar c$ production cross-section below the open charm threshold to individual charmonium states. Comparisons of ICEM predictions with the parameter values and their uncertainties previously determined from fits to LHCb data at 7~\TeV~\cite{LHCb7tev_1,Aaij:2014qea} are shown in the bottom panels of Figures~\ref{fig:Pred_Jpsi_P}(a) and~\ref{fig:Pred_Jpsi_P}(b) for $\jpsi$ and $\psip$ respectively.
The model seems to predict somewhat harder $\pt$ spectra than observed in the data for both $\jpsi$ and $\psip$, and tends to underestimate the cross-section for $\psip$.

\subsection{Theory comparison: non-prompt production}
 
\begin{figure}[htb]
\subfigure[]{\includegraphics[width=.49\columnwidth]{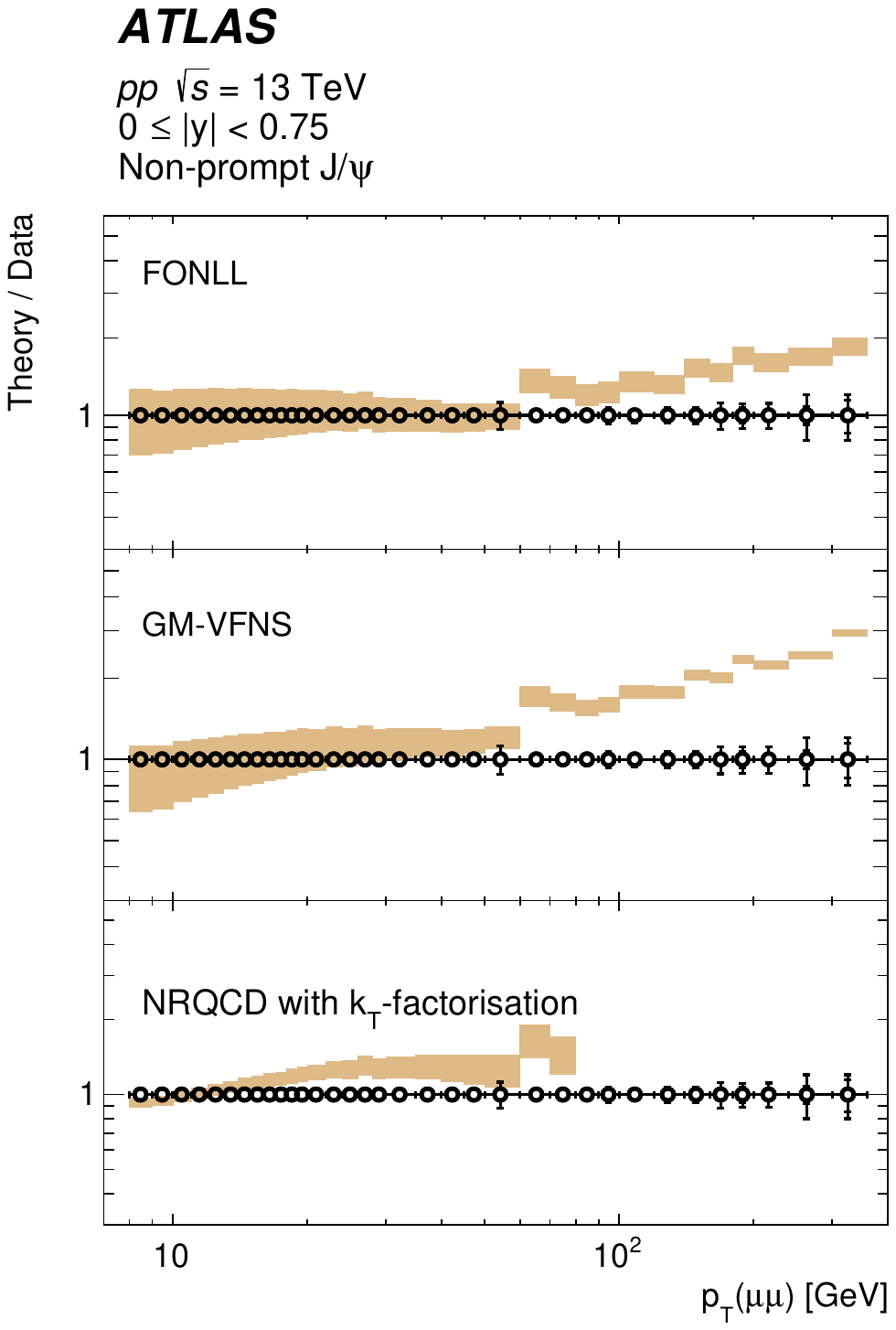}}
\subfigure[]{\includegraphics[width=.49\columnwidth]{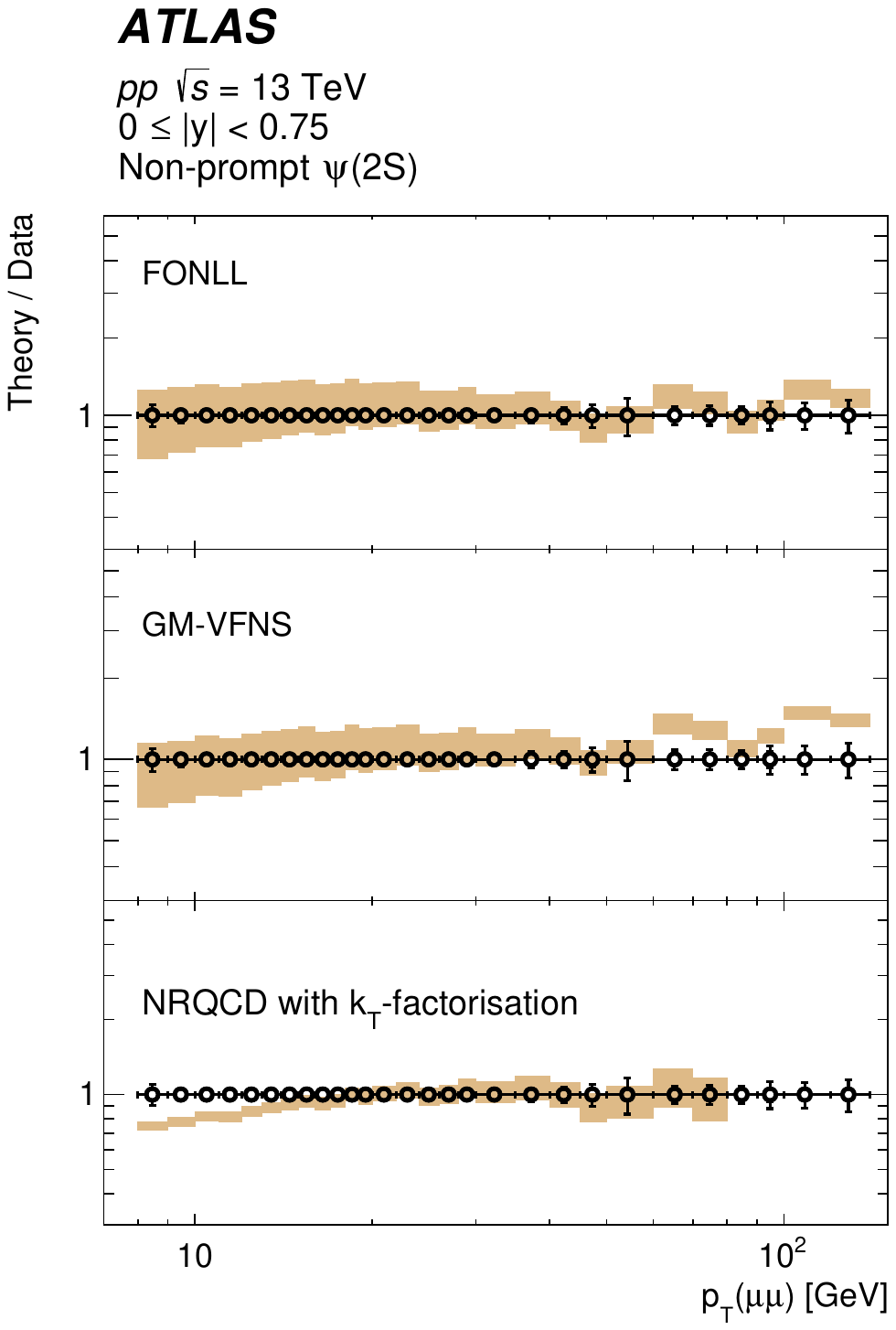}}
\caption{Ratios of various theoretical predictions (described in the text) to the data points from this measurement, for non-prompt production of (a) $\jpsi$ and (b)  $\psip$ in the central rapidity region. In each $\pt$ bin, the shaded area represents the ratio of the theoretical prediction to the measured value, with the vertical spread showing the uncertainties of the respective model.
Error bars on the black circles show fractional uncertainties of this measurement.
}
\label{fig:Pred_Jpsi_NP}
\end{figure}

Theoretical calculations of non-prompt charmonium production are based on perturbative QCD for the production of a $b \bar b$ quark pair, their hadronisation into a pair of $b$-hadrons, and their subsequent decay into a charmonium state with specific quantum numbers. Predictions from one such model, based on fixed-order--next-to-leading-log (FONLL) QCD calculations~\cite{FONLL_2001,FONLL_2012} were obtained using the web-based tool~\cite{FONLL_WEB} with default parameter values, and are shown in comparison with our measurements in the top panels of Figures~\ref{fig:Pred_Jpsi_NP}(a) and~\ref{fig:Pred_Jpsi_NP}(b) for $\jpsi$ and $\psip$ respectively. Here the uncertainties cover variations of both the renormalisation scale and the charm quark mass. Agreement is good at lower~$\pt$ values, but the FONLL model predicts higher cross-sections for $\jpsi$ at the high $\pt$ end.

Another set of predictions, based on the next-to-leading-order QCD calculation in the general-mass--variable-flavour-number scheme (GM--VFNS)~\cite{Kram_1} are shown in the middle panels of Figures~\ref{fig:Pred_Jpsi_NP}(a) and~\ref{fig:Pred_Jpsi_NP}(b).
Parameter values were determined in Ref.~\cite{PRD842011051501,Kram_2}, with uncertainties originating from renormalisation scale dependence. These predictions lead to similar results, but the deviation from data at the highest $\pt$ values is somewhat more pronounced, especially in the $\jpsi$ case.
 
Finally, the NRQCD model with $k_T$-factorisation can also be used to predict the $\pt$ distributions of vector charmonia through the non-prompt production mechanisms~\cite{Eur.Phys.J.C788202018,Eur.Phys.J.C803302020} (see bottom panels of Figures~\ref{fig:Pred_Jpsi_NP}(a) and~\ref{fig:Pred_Jpsi_NP}(b)). Where available, the shapes of $\pt$ distributions are reproduced fairly well for both $\jpsi$ and $\psip$, but the limitations of the transverse-momentum-dependent model for the gluon PDF show up at even lower charmonium $\pt$ values. Also, in this model the cross-section for $\psip$ non-prompt production at low $\pt$ is somewhat underestimated.
 
Overall, none of the models considered here is able to describe the data  over the whole measured range of transverse momenta.
The general trend shown by all theoretical models is a slower-than-observed decrease of the cross-section with $\pt$, which could be related to insufficiently accounting for PDF evolution and/or possible dependence of LDMEs on transverse momentum. In any case, these measurements should help refine theoretical models of hadronic production of quarkonium at the highest available energies and at transverse momenta well beyond 100~\GeV.

% End of text imported from the .//Results.tex input file

\FloatBarrier
 
\section{Summary}
\label{sec:conclusion}

% The next lines are included from the .//Summary.tex input file
 
This paper describes a measurement of the double-differential production cross-sections of $\jpsi$ and $\psip$ charmonium states
in $pp$ collisions at $\sqrt{s}=13~{\si{\tev}}$, performed through their decays into dimuons and using 140~fb$^{-1}$ of data collected by the
ATLAS detector at the LHC during \RunTwo.
The cross-sections for each of the two states are measured separately for prompt and non-prompt production mechanisms.
The non-prompt fractions
for each state are also measured, along with the $\psip$-to-$\jpsi$ production ratios. The rapidity range of the measurement is $|y|<2$.
For $\psip$ meson the transverse momentum range is $8{-}140~{\si{\gev}}$, while for $\jpsi$ state the results cover a much wider transverse momentum range, from $8~{\si{\gev}}$ to $360~{\si{\gev}}$, extending well beyond the range of previous measurements.
In the high $\pt$ range the results show similar $\pt$-dependences for the prompt and non-prompt differential cross-sections, with the non-prompt fractions being nearly constant for both $\jpsi$ and $\psip$ states.
 
The results are compared with a number of theoretical predictions, which describe the data with varying degrees of success. The extended $\pt$ reach of this measurement provides important fresh input  for future tuning of theoretical models.

% End of text imported from the .//Summary.tex input file

\section*{Acknowledgements}

% The next lines are included from the .//acknowledgements/Acknowledgements.tex input file

We thank CERN for the very successful operation of the LHC, as well as the
support staff from our institutions without whom ATLAS could not be
operated efficiently.
 
We acknowledge the support of
ANPCyT, Argentina;
YerPhI, Armenia;
ARC, Australia;
BMWFW and FWF, Austria;
ANAS, Azerbaijan;
CNPq and FAPESP, Brazil;
NSERC, NRC and CFI, Canada;
CERN;
ANID, Chile;
CAS, MOST and NSFC, China;
Minciencias, Colombia;
MEYS CR, Czech Republic;
DNRF and DNSRC, Denmark;
IN2P3-CNRS and CEA-DRF/IRFU, France;
SRNSFG, Georgia;
BMBF, HGF and MPG, Germany;
GSRI, Greece;
RGC and Hong Kong SAR, China;
ISF and Benoziyo Center, Israel;
INFN, Italy;
MEXT and JSPS, Japan;
CNRST, Morocco;
NWO, Netherlands;
RCN, Norway;
MEiN, Poland;
FCT, Portugal;
MNE/IFA, Romania;
MESTD, Serbia;
MSSR, Slovakia;
ARRS and MIZ\v{S}, Slovenia;
DSI/NRF, South Africa;
MICINN, Spain;
SRC and Wallenberg Foundation, Sweden;
SERI, SNSF and Cantons of Bern and Geneva, Switzerland;
MOST, Taipei;
TENMAK, T\"urkiye;
STFC, United Kingdom;
DOE and NSF, United States of America.
In addition, individual groups and members have received support from
BCKDF, CANARIE, CRC and DRAC, Canada;
PRIMUS 21/SCI/017 and UNCE SCI/013, Czech Republic;
COST, ERC, ERDF, Horizon 2020, ICSC-NextGenerationEU and Marie Sk{\l}odowska-Curie Actions, European Union;
Investissements d'Avenir Labex, Investissements d'Avenir Idex and ANR, France;
DFG and AvH Foundation, Germany;
Herakleitos, Thales and Aristeia programmes co-financed by EU-ESF and the Greek NSRF, Greece;
BSF-NSF and MINERVA, Israel;
Norwegian Financial Mechanism 2014-2021, Norway;
NCN and NAWA, Poland;
La Caixa Banking Foundation, CERCA Programme Generalitat de Catalunya and PROMETEO and GenT Programmes Generalitat Valenciana, Spain;
G\"{o}ran Gustafssons Stiftelse, Sweden;
The Royal Society and Leverhulme Trust, United Kingdom.
 
The crucial computing support from all WLCG partners is acknowledged gratefully, in particular from CERN, the ATLAS Tier-1 facilities at TRIUMF/SFU (Canada), NDGF (Denmark, Norway, Sweden), CC-IN2P3 (France), KIT/GridKA (Germany), INFN-CNAF (Italy), NL-T1 (Netherlands), PIC (Spain), RAL (UK) and BNL (USA), the Tier-2 facilities worldwide and large non-WLCG resource providers. Major contributors of computing resources are listed in Ref.~\cite{ATL-SOFT-PUB-2023-001}.

% End of text imported from the .//acknowledgements/Acknowledgements.tex input file

\appendix
\part*{Appendix}
\label{sec:spinal}

% The next lines are included from the .//SpinAlignment.tex input file

The $\psi$ meson's acceptance depends on its spin alignment. The nominal results presented in this paper assume an isotropic angular distribution, corresponding to the \enquote{unpolarised} case.
In order to assess the dependence of the cross-section on the spin alignment, the acceptance maps were reweighted for a variety of spin-alignment scenarios, and for each analysis bin the correction factors were determined relative to the nominal values.
The correction factors calculated for
several values of the parameter $\lambda_{\theta}$, assuming the angular dependence in the $\psi$ helicity frame to be $\propto(1+\lambda_{\theta} \cos^2 \theta)$, are shown in Table~\ref{tbl:spin_long}.
The factors are shown for prompt $\jpsi$ in the central rapidity range,
but were found to be the same for $\jpsi$ and $\psip$ within 1\%--2\%, with the variation over the other  rapidity ranges also within 1\%--2\%.
For $\lambda_{\theta} \ne 0$
the nominal value of the measured cross-section should be multiplied by the appropriate correction factor from Table~\ref{tbl:spin_long}.

\begin{table}[H]
\begin{center}
\caption{Correction factors for an
assumed angular dependence $\propto 1+\lambda_{\theta} \cos^2\theta $ in the helicity frame, for several values of the parameter $\lambda_{\theta}$. The correction factors were found to be the same (within 1\%--2\%) for $\jpsi$ and $\psip$ mesons, for prompt and non-prompt production mechanisms, and for the three rapidity intervals considered in this paper.}
\label{tbl:spin_long}
\begin{tabular}{|c|c|c|c|c|}
\hline
\pt & $\lambda_{\theta} = -1$ & $\lambda_{\theta} = -0.2$ & $\lambda_{\theta} = +0.2$ & $\lambda_{\theta} = +1$ \\
\hline
\phantom{10}8--\phantom{10}9 [GeV]   & 0.69 & 0.93 & 1.07 & 1.29 \\
\phantom{10}9--\phantom{1}10 [GeV]  & 0.69 & 0.94 & 1.06 & 1.28 \\
\phantom{1}10--\phantom{1}11 [GeV] & 0.70 & 0.94 & 1.06 & 1.27 \\
\phantom{1}11--\phantom{1}12 [GeV] & 0.70 & 0.94 & 1.06 & 1.26 \\
\phantom{1}12--\phantom{1}13 [GeV] & 0.71 & 0.94 & 1.06 & 1.25 \\
\phantom{1}13--\phantom{1}14 [GeV] & 0.72 & 0.94 & 1.05 & 1.24 \\
\phantom{1}14--\phantom{1}15 [GeV] & 0.73 & 0.95 & 1.05 & 1.24 \\
\phantom{1}15--\phantom{1}16 [GeV] & 0.73 & 0.95 & 1.05 & 1.22 \\
\phantom{1}16--\phantom{1}17 [GeV] & 0.74 & 0.95 & 1.05 & 1.21 \\
\phantom{1}17--\phantom{1}18 [GeV] & 0.75 & 0.95 & 1.05 & 1.20 \\
\phantom{1}18--\phantom{1}19 [GeV] & 0.75 & 0.95 & 1.04 & 1.20 \\
\phantom{1}19--\phantom{1}20 [GeV] & 0.76 & 0.96 & 1.04 & 1.19 \\
\phantom{1}20--\phantom{1}22 [GeV] & 0.77 & 0.96 & 1.04 & 1.18 \\
\phantom{1}22--\phantom{1}24 [GeV] & 0.78 & 0.96 & 1.04 & 1.17 \\
\phantom{1}24--\phantom{1}26 [GeV] & 0.79 & 0.96 & 1.03 & 1.15 \\
\phantom{1}26--\phantom{1}28 [GeV] & 0.80 & 0.97 & 1.03 & 1.14 \\
\phantom{1}28--\phantom{1}30 [GeV] & 0.81 & 0.97 & 1.03 & 1.13 \\
\phantom{1}30--\phantom{1}35 [GeV] & 0.82 & 0.97 & 1.03 & 1.12 \\
\phantom{1}35--\phantom{1}40 [GeV] & 0.84 & 0.97 & 1.02 & 1.10 \\
\phantom{1}40--\phantom{1}45 [GeV] & 0.86 & 0.98 & 1.02 & 1.09 \\
\phantom{1}45--\phantom{1}50 [GeV] & 0.87 & 0.98 & 1.02 & 1.08 \\
\phantom{1}50--\phantom{1}60 [GeV] & 0.88 & 0.98 & 1.02 & 1.07 \\
\hline
\phantom{1}60--\phantom{1}70 [GeV] & 1.48 & 1.05 & 0.96 & 0.86 \\
\phantom{1}70--\phantom{1}80 [GeV] & 1.19 & 1.02 & 0.98 & 0.93 \\
\phantom{1}80--\phantom{1}90 [GeV] & 1.05 & 1.01 & 0.99 & 0.98 \\
\phantom{1}90--100 [GeV] & 0.98 & 1.00 & 1.00 & 1.01 \\
100--120 [GeV] & 0.94 & 0.99 & 1.01 & 1.03 \\
120--140 [GeV] & 0.94 & 0.99 & 1.01 & 1.03 \\
140--160 [GeV] & 0.95 & 0.99 & 1.01 & 1.03 \\
160--180 [GeV] & 0.96 & 0.99 & 1.01 & 1.02 \\
180--200 [GeV] & 0.96 & 0.99 & 1.00 & 1.02 \\
200--240 [GeV] & 0.97 & 1.00 & 1.00 & 1.02 \\
240--300 [GeV] & 0.97 & 1.00 & 1.00 & 1.01 \\
300--360 [GeV] & 0.98 & 1.00 & 1.00 & 1.01 \\
\hline
\end{tabular}
\end{center}
\end{table}

% End of text imported from the .//SpinAlignment.tex input file

\printbibliography

@article{LUCID2,
  author={G. Avoni and others},
  title={The new LUCID-2 detector for luminosity measurement and monitoring in ATLAS},
  journal={JINST},
  volume={13},
  number={07},
  pages={P07017},
  doi="10.1088/1748-0221/13/07/P07017",
  year={2018},
}

@article{Aaij:2014qea,
      author         = "{LHCb Collaboration}",
      title          = "{Measurement of $\psi(2S)$ polarisation in $pp$
                        collisions at $\sqrt{s}=7$ TeV}",
      collaboration  = "LHCb Collaboration",
      journal        = "Eur. Phys. J. C",
      volume         = "74",
      year           = "2014",
      number         = "5",
      pages          = "2872",
      doi            = "10.1140/epjc/s10052-014-2872-9",
      eprint         = "1403.1339",
      archivePrefix  = "arXiv",
      primaryClass   = "hep-ex",
      reportNumber   = "LHCB-PAPER-2013-067, CERN-PH-EP-2014-029",
      SLACcitation   = "%%CITATION = ARXIV:1403.1339;%%"
}

@article{Abelev:2011md,
      author         = "{ALICE Collaboration}",
      title          = "{$J/\psi$ polarization in $pp$ collisions at $\sqrt{s}=7$
                        TeV}",
      collaboration  = "ALICE Collaboration",
      journal        = "Phys. Rev. Lett.",
      volume         = "108",
      year           = "2012",
      pages          = "082001",
      doi            = "10.1103/PhysRevLett.108.082001",
      eprint         = "1111.1630",
      archivePrefix  = "arXiv",
      primaryClass   = "hep-ex",
      reportNumber   = "CERN-PH-EP-2011-182",
      SLACcitation   = "%%CITATION = ARXIV:1111.1630;%%"
}

@article{Acharya:2018uww,
      author         = "{ALICE Collaboration}",
      title          = "{Measurement of the inclusive J/$\psi$ polarization at
                        forward rapidity in pp collisions at $\mathbf {\sqrt{s} =
                        8}$  TeV}",
      collaboration  = "ALICE Collaboration",
      journal        = "Eur. Phys. J. C",
      volume         = "78",
      year           = "2018",
      number         = "7",
      pages          = "562",
      doi            = "10.1140/epjc/s10052-018-6027-2",
      eprint         = "1805.04374",
      archivePrefix  = "arXiv",
      primaryClass   = "hep-ex",
      reportNumber   = "CERN-EP-2018-098",
      SLACcitation   = "%%CITATION = ARXIV:1805.04374;%%"
}

@article{FONLL_2001,
    author = "Cacciari, Matteo and Frixione, Stefano and Nason, Paolo",
    title = "{The $p_T$ spectrum in heavy flavor photoproduction}",
    eprint = "hep-ph/0102134",
    archivePrefix = "arXiv",
    reportNumber = "BICOCCA-FT-01-01, GEF-TH-2-01, YITP-SB-01-01",
    doi = "10.1088/1126-6708/2001/03/006",
    journal = "JHEP",
    volume = "03",
    pages = "006",
    year = "2001"
}

@article{FONLL_2012,
    author = "Cacciari, Matteo and Frixione, Stefano and Houdeau, Nicolas and Mangano, Michelangelo L. and Nason, Paolo and Ridolfi, Giovanni",
    title = "{Theoretical predictions for charm and bottom production at the LHC}",
    eprint = "1205.6344",
    archivePrefix = "arXiv",
    primaryClass = "hep-ph",
    reportNumber = "CERN-PH-TH-2011-227",
    doi = "10.1007/JHEP10(2012)137",
    journal = "JHEP",
    volume = "10",
    pages = "137",
    year = "2012"
}

@article{Bodwin:1994jh,
      author         = "Bodwin, Geoffrey T. and Braaten, Eric and Lepage, G.
                        Peter",
      title          = "{Rigorous QCD analysis of inclusive annihilation and
                        production of heavy quarkonium}",
      journal        = "Phys. Rev. D",
      volume         = "51",
      year           = "1995",
      pages          = "1125-1171",
      doi            = "10.1103/PhysRevD.51.1125",
      addendum       = "[Erratum: \href{https://doi.org/10.1103/PhysRevD.55.5853}{Phys. Rev. D \textbf{55} (1997)}]",      
      eprint         = "hep-ph/9407339",
      archivePrefix  = "arXiv",
      primaryClass   = "hep-ph",
      reportNumber   = "ANL-HEP-PR-94-24, FERMILAB-PUB-94-073-T, NUHEP-TH-94-5",
      SLACcitation   = "%%CITATION = HEP-PH/9407339;%%"
}

@article{Aad:2014rua,
      author         = "{ATLAS Collaboration}",
      title          = "{Measurement of the production cross section of prompt
                        $J/\psi$ mesons in association with a $W^\pm$ boson in
                        $pp$ collisions at $\sqrt{s} =$ 7 TeV with the ATLAS
                        detector}",
      collaboration  = "{ATLAS Collaboration}",
      journal        = "JHEP",
      volume         = "04",
      year           = "2014",
      pages          = "172",
      doi            = "10.1007/JHEP04(2014)172",
      eprint         = "1401.2831",
      archivePrefix  = "arXiv",
      primaryClass   = "hep-ex",
      reportNumber   = "CERN-PH-EP-2013-184",
      SLACcitation   = "%%CITATION = ARXIV:1401.2831;%%"
}

@article{Aad:2014kba,
      author         = "{ATLAS Collaboration}",
      title          = "{Observation and measurements of the production of prompt
                        and non-prompt $J/\psi$ mesons in association with a $Z$
                        boson in $pp$ collisions at $\sqrt{s}$ = 8 TeV with the
                        ATLAS detector}",
      collaboration  = "ATLAS",
      journal        = "Eur. Phys. J. C",
      volume         = "75",
      year           = "2015",
      number         = "5",
      pages          = "229",
      doi            = "10.1140/epjc/s10052-015-3406-9",
      eprint         = "1412.6428",
      archivePrefix  = "arXiv",
      primaryClass   = "hep-ex",
      reportNumber   = "CERN-PH-EP-2014-276",
      SLACcitation   = "%%CITATION = ARXIV:1412.6428;%%"
}

@article{Li:2010hc,
      author         = "Li, Gang and Song, Mao and Zhang, Ren-You and Ma,
                        Wen-Gan",
      title          = "{QCD corrections to $J/\psi$ production in association
                        with a $W$-boson at the LHC}",
      journal        = "Phys. Rev. D",
      volume         = "83",
      year           = "2011",
      pages          = "014001",
      doi            = "10.1103/PhysRevD.83.014001",
      eprint         = "1012.3798",
      archivePrefix  = "arXiv",
      primaryClass   = "hep-ph",
      SLACcitation   = "%%CITATION = ARXIV:1012.3798;%%"
}

@article{Lansberg:2013wva,
      author         = "Lansberg, J. P. and Lorce, C.",
      title          = "{Reassessing the importance of the colour-singlet
                        contributions to direct $J/\psi + W$ production at the LHC
                        and the Tevatron}",
      journal        = "Phys. Lett. B",
      volume         = "726",
      year           = "2013",
      pages          = "218-222",
      doi            = "10.1016/j.physletb.2013.07.059",
                        
      addendum       = "[Erratum: \href{https://doi.org/10.1016/j.physletb.2014.10.015}{Phys. Lett. B \textbf{738} (2014) 529}]",  
      eprint         = "1303.5327",
      archivePrefix  = "arXiv",
      primaryClass   = "hep-ph",
      SLACcitation   = "%%CITATION = ARXIV:1303.5327;%%"
}

@article{Gong:2012ah,
      author         = "Gong, Bin and Lansberg, Jean-Philippe and Lorce, Cedric
                        and Wang, Jianxiong",
      title          = "{Next-to-leading-order QCD corrections to the yields and
                        polarisations of $J/\psi$ and $\Upsilon$ directly produced in
                        association with a Z boson at the LHC}",
      journal        = "JHEP",
      volume         = "03",
      year           = "2013",
      pages          = "115",
      doi            = "10.1007/JHEP03(2013)115",
      eprint         = "1210.2430",
      archivePrefix  = "arXiv",
      primaryClass   = "hep-ph",
      SLACcitation   = "%%CITATION = ARXIV:1210.2430;%%"
}

@article{Mao:2011kf,
      author         = "Song, Mao and Ma, Wen-Gan and Li, Gang and Zhang, Ren-You
                        and Guo, Lei",
      title          = "{QCD corrections to $J/\psi$ plus $Z^0$-boson production
                        at the LHC}",
      journal        = "JHEP",
      volume         = "02",
      year           = "2011",
      pages          = "071",
      doi            = "10.1007/JHEP02(2011)071",
      addendum       = "[Erratum: \href{https://doi.org/10.1007/JHEP12(2012)010}{JHEP \textbf{12} (2012) 010}]", 
      eprint         = "1102.0398",
      archivePrefix  = "arXiv",
      primaryClass   = "hep-ph",
      SLACcitation   = "%%CITATION = ARXIV:1102.0398;%%"
}

@article{Butenschoen:2012qh,
      author         = "Butenschoen, Mathias and Kniehl, Bernd A.",
      title          = "{J/$\psi$ production in NRQCD: A global analysis of yield
                        and polarization}",
      booktitle      = "{Proceedings, Ringberg Workshop on New trends in HERA
                        physics 2011: Ringberg Castle, Tegernsee, Germany,
                        September 25-28, 2011}",
      journal        = "Nucl. Phys. Proc. Suppl.",
      volume         = "222-224",
      year           = "2012",
      pages          = "151-161",
      doi            = "10.1016/j.nuclphysbps.2012.03.016",
      eprint         = "1201.3862",
      archivePrefix  = "arXiv",
      primaryClass   = "hep-ph",
      SLACcitation   = "%%CITATION = ARXIV:1201.3862;%%"
}

@article{Chatrchyan:2011kc,
      author         = "{CMS Collaboration}",
      title          = "{$J/\psi$ and $\psi(2S)$ production in $pp$ collisions
                        at $\sqrt{s}=7$ TeV}",
      collaboration  = "CMS",
      journal        = "JHEP",
      volume         = "02",
      year           = "2012",
      pages          = "011",
      doi            = "10.1007/JHEP02(2012)011",
      eprint         = "1111.1557",
      archivePrefix  = "arXiv",
      primaryClass   = "hep-ex",
      reportNumber   = "CMS-BPH-10-014, CERN-PH-EP-2011-177",
      SLACcitation   = "%%CITATION = ARXIV:1111.1557;%%"
}

@article{Aaij:2012ag,
      author         = "{LHCb Collaboration}",
      title          = "{Measurement of $\psi(2S)$ meson production in $pp$
                        collisions at $\sqrt{s}$=7 TeV}",
      collaboration  = "LHCb",
      journal        = "Eur. Phys. J. C",
      volume         = "72",
      year           = "2012",
      pages          = "2100",
      doi            = "10.1140/epjc/s10052-012-2100-4",
      eprint         = "1204.1258",
      archivePrefix  = "arXiv",
      primaryClass   = "hep-ex",
      reportNumber   = "CERN-PH-EP-2012-094, LHCB-PAPER-2011-045",
      SLACcitation   = "%%CITATION = ARXIV:1204.1258;%%"
}

@article{Abelev:2014qha,
      author         = "{ALICE Collaboration}",
      title          = "{Measurement of quarkonium production at forward rapidity
                        in $pp$ collisions at $\sqrt{s} = 7$ TeV}",
      collaboration  = "ALICE",
      journal        = "Eur. Phys. J. C",
      volume         = "74",
      year           = "2014",
      number         = "8",
      pages          = "2974",
      doi            = "10.1140/epjc/s10052-014-2974-4",
      eprint         = "1403.3648",
      archivePrefix  = "arXiv",
      primaryClass   = "nucl-ex",
      reportNumber   = "CERN-PH-EP-2014-042",
      SLACcitation   = "%%CITATION = ARXIV:1403.3648;%%"
}

@misc{FONLL_WEB,
      author = "M.~Cacciari",
      title = {{FONLL Heavy Quark Production}},
      howpublished = {\url{http://www.lpthe.jussieu.fr/~cacciari/fonll/fonllform.html}},
      note = {Accessed: 2019-09-03}
}

@article{ALICE2021,
    author         = "{ALICE Collaboration}",
      %author = "Acharya, Shreyasi and others",
    collaboration = "ALICE",
    title = "{Prompt and non-prompt J/\ensuremath{\psi} production cross sections at midrapidity in proton-proton collisions at $ \sqrt{\mathrm{s}} $ = 5.02 and 13 TeV}",
    eprint = "2108.02523",
    archivePrefix = "arXiv",
    primaryClass = "nucl-ex",
    reportNumber = "CERN-EP-2021-152",
    doi = "10.1007/JHEP03(2022)190",
    journal = "JHEP",
    volume = "03",
    pages = "190",
    year = "2022"
}

@article{PRL1062011022003,
  title = {Reconciling $J/\ensuremath{\psi}$ Production at HERA, RHIC, Tevatron, and LHC with Nonrelativistic QCD Factorization at Next-to-Leading Order},
  author = {Butensch\"on, Mathias and Kniehl, Bernd A.},
  journal = {Phys. Rev. Lett.},
  volume = {106},
  pages = {022003},
  numpages = {4},
  year = {2011},
  month = {Jan},
  publisher = {American Physical Society},
  doi = {10.1103/PhysRevLett.106.022003}%,
}

@article{PRD842011051501,
  title = {World data of $J/\ensuremath{\psi}$ production consolidate nonrelativistic QCD factorization at next-to-leading order},
  author = {Butenschoen, Mathias and Kniehl, Bernd A.},
  journal = {Phys. Rev. D},
  volume = {84},
  pages = {051501},
  numpages = {5},
  year = {2011},
  month = {Sep},
  publisher = {American Physical Society},
  doi = {10.1103/PhysRevD.84.051501}%,
}

@article{BaranovEur.Phys.J.C754552015,
 author = "Baranov, S. P. and Lipatov, A. V. and Zotov, N. P.",
    title = "{Prompt charmonia production and polarization at LHC in the NRQCD with $k_T$ -factorization. Part I: $\psi (2S)$ meson}",
    eprint = "1508.05480",
    archivePrefix = "arXiv",
    primaryClass = "hep-ph",
    reportNumber = "DESY-15-177",
    doi = "10.1140/epjc/s10052-015-3689-x",
    journal = "Eur. Phys. J. C",
    volume = "75",
    number = "9",
    pages = "455",
    year = "2015"
}

@article{Phys.Rev.D960340192017,
  title = {Prompt charmonia production and polarization at the LHC in the NRQCD with ${k}_{T}$-factorization. III. $J/\ensuremath{\psi}$ meson},
  author = {Baranov, S. P. and Lipatov, A. V.},
  journal = {Phys. Rev. D},
  volume = {96},
  pages = {034019},
  numpages = {13},
  year = {2017},
  month = {Aug},
  publisher = {American Physical Society},
  doi = {10.1103/PhysRevD.96.034019}%,
}

@article{Eur.Phys.J.C788202018,
    author = "Baranov, S. P. and Lipatov, A. V. and Malyshev, M. A.",
    title = "{Associated non-prompt $J/\psi + \mu$ and $J/\psi + J/\psi$ production at LHC as a test for TMD gluon density}",
    eprint = "1808.06233",
    archivePrefix = "arXiv",
    primaryClass = "hep-ph",
    reportNumber = "DESY-18-138, DESY 18-138",
    doi = "10.1140/epjc/s10052-018-6297-8",
    journal = "Eur. Phys. J. C",
    volume = "78",
    number = "10",
    pages = "820",
    year = "2018"
}

@article{Eur.Phys.J.C803302020,
    author = "Lipatov, A. V. and Malyshev, M. A. and Baranov, S. P.",
    title = "{Particle Event Generator: A Simple-in-Use System PEGASUS version 1.0}",
    eprint = "1912.04204",
    archivePrefix = "arXiv",
    primaryClass = "hep-ph",
    doi = "10.1140/epjc/s10052-020-7898-6",
    journal = "Eur. Phys. J. C",
    volume = "80",
    number = "4",
    pages = "330",
    year = "2020"
}

@article{Phys.Rev.D1001140212019,
  title = {Are there any challenges in the charmonia production and polarization at the LHC?},
  author = {Baranov, S. P. and Lipatov, A. V.},
  journal = {Phys. Rev. D},
  volume = {100},
  pages = {114021},
  numpages = {19},
  year = {2019},
  month = {Dec},
  publisher = {American Physical Society},
  doi = {10.1103/PhysRevD.100.114021}%,
}

@article{PhysRevD.104.094026,
  title = {Production and polarization of direct $J/\ensuremath{\psi}$ to $\mathcal{O}({\ensuremath{\alpha}}_{s}^{3})$ in the improved color evaporation model in collinear factorization},
  author = {Cheung, Vincent and Vogt, Ramona},
  journal = {Phys. Rev. D},
  volume = {104},
  pages = {094026},
  numpages = {11},
  year = {2021},
  month = {Nov},
  publisher = {American Physical Society},
  doi = {10.1103/PhysRevD.104.094026}%,
}

@article{LHCb7tev_1,
    author         = "{LHCb Collaboration}",
    collaboration = "LHCb",
    title = "{Measurement of $J/\psi$ production in $pp$ collisions at $\sqrt{s}=7$ TeV}",
    eprint = "1103.0423",
    archivePrefix = "arXiv",
    primaryClass = "hep-ex",
    reportNumber = "LHCb-PAPER-2011-003, CERN-PH-EP-2011-018",
    doi = "10.1140/epjc/s10052-011-1645-y",
    journal = "Eur. Phys. J. C",
    volume = "71",
    pages = "1645",
    year = "2011"
}

@article{Kram_1,
  title = {Inclusive $J/\ensuremath{\psi}$ and $\ensuremath{\psi}(2S)$ production from $b$-hadron decay in $p\overline{p}$ and $pp$ collisions},
  author = {Bolzoni, Paolo and Kniehl, Bernd A. and Kramer, Gustav},
  journal = {Phys. Rev. D},
  volume = {88},
  pages = {074035},
  numpages = {6},
  year = {2013},
  month = {Oct},
  publisher = {American Physical Society},
  doi = {10.1103/PhysRevD.88.074035}%,
}

@article{Kram_2,
  title = {Cross sections of inclusive $\ensuremath{\psi}(2S)$ and $X(3872)$ production from $b$-hadron decays in $pp$ collisions and comparison with ATLAS, CMS, and LHCb data},
  author = {Kniehl, B. A. and Kramer, G. and Schienbein, I. and Spiesberger, H.},
  journal = {Phys. Rev. D},
  volume = {103},
  pages = {094002},
  numpages = {6},
  year = {2021},
  month = {May},
  publisher = {American Physical Society},
  doi = {10.1103/PhysRevD.103.094002}%,
}

@article{LHCb2,
    author         = "{LHCb Collaboration}",
    collaboration = "LHCb",
    title = "{Exclusive $J/\psi$ and $\psi$(2S) production in pp collisions at $ \sqrt{s} = 7$ TeV}",
    eprint = "1301.7084",
    archivePrefix = "arXiv",
    primaryClass = "hep-ex",
    reportNumber = "CERN-PH-EP-2013-005, LHCB-PAPER-2012-044",
    doi = "10.1088/0954-3899/40/4/045001",
    journal = "J. Phys. G",
    volume = "40",
    pages = "045001",
    year = "2013"
}

@article{LHCb3,
    author         = "{LHCb Collaboration}",
    collaboration = "LHCb",
    title = "{Measurement of $J/\psi$ polarization in $pp$ collisions at $\sqrt{s}=7$ TeV}",
    eprint = "1307.6379",
    archivePrefix = "arXiv",
    primaryClass = "hep-ex",
    reportNumber = "LHCB-PAPER-2013-008, CERN-PH-EP-2013-130",
    doi = "10.1140/epjc/s10052-013-2631-3",
    journal = "Eur. Phys. J. C",
    volume = "73",
    number = "11",
    pages = "2631",
    year = "2013"
}

@article{LHCb4,
    author         = "{LHCb Collaboration}",
    collaboration = "LHCb",
    title = "{Updated measurements of exclusive $J/\psi$ and $\psi$(2S) production cross-sections in pp collisions at $\sqrt{s}=7$ TeV}",
    eprint = "1401.3288",
    archivePrefix = "arXiv",
    primaryClass = "hep-ex",
    reportNumber = "CERN-PH-EP-2013-233, LHCB-PAPER-2013-059",
    doi = "10.1088/0954-3899/41/5/055002",
    journal = "J. Phys. G",
    volume = "41",
    pages = "055002",
    year = "2014"
}

@article{LHCb5,
    author         = "{LHCb Collaboration}",
    title = "{Measurement of forward $J/\psi$ production cross-sections in $pp$ collisions at $\sqrt{s}=13$ TeV}",
    eprint = "1509.00771",
    archivePrefix = "arXiv",
    primaryClass = "hep-ex",
    reportNumber = "LHCB-PAPER-2015-037, CERN-PH-EP-2015-222",
    doi = "10.1007/JHEP10(2015)172",
    journal = "JHEP",
    volume = "10",
    pages = "172",
    year = "2015",
    addendum       = "[Erratum: \href{https://link.springer.com/article/10.1007/JHEP05(2017)063}{JHEP \textbf{05} (2017) 063}]",
}

@article{LHCb6,
    author         = "{LHCb Collaboration}",
    collaboration = "LHCb",
    title = "{Central exclusive production of $J/\psi$ and $\psi(2S)$ mesons in $pp$ collisions at $\sqrt{s}=13~$TeV}",
    eprint = "1806.04079",
    archivePrefix = "arXiv",
    primaryClass = "hep-ex",
    reportNumber = "LHCB-PAPER-2018-011, LHCb-PAPER-2018-011, CERN-EP-2018-152",
    doi = "10.1007/JHEP10(2018)167",
    journal = "JHEP",
    volume = "10",
    pages = "167",
    year = "2018"
}

@article{Butenschoen:2022orc,
    author = "Butenschoen, Mathias and Kniehl, Bernd A.",
    title = "{Global analysis of \ensuremath{\psi}(2S) inclusive hadroproduction at next-to-leading order in nonrelativistic-QCD factorization}",
    eprint = "2207.09346",
    archivePrefix = "arXiv",
    primaryClass = "hep-ph",
    reportNumber = "DESY-22-119",
    doi = "10.1103/PhysRevD.107.034003",
    journal = "Phys. Rev. D",
    volume = "107",
    number = "3",
    pages = "034003",
    year = "2023"
}

@article{ALICE1,
    author         = "{ALICE Collaboration}",
    collaboration = "ALICE",
    title = "{Prompt and non-prompt J/\ensuremath{\psi} production cross sections at midrapidity in proton-proton collisions at $ \sqrt{\mathrm{s}} $ = 5.02 and 13 TeV}",
    eprint = "2108.02523",
    archivePrefix = "arXiv",
    primaryClass = "nucl-ex",
    reportNumber = "CERN-EP-2021-152",
    doi = "10.1007/JHEP03(2022)190",
    journal = "JHEP",
    volume = "03",
    pages = "190",
    year = "2022"
}

@article{ALICE2,
    author         = "{ALICE Collaboration}",
    collaboration = "ALICE",
    title = "{Inclusive $\text {J}/\psi $ production at midrapidity in pp collisions at $\sqrt{s} = 13$ TeV}",
    eprint = "2108.01906",
    archivePrefix = "arXiv",
    primaryClass = "nucl-ex",
    reportNumber = "CERN-EP-2021-160",
    doi = "10.1140/epjc/s10052-021-09873-4",
    journal = "Eur. Phys. J. C",
    volume = "81",
    number = "12",
    pages = "1121",
    year = "2021"
}

@article{ALICE4,
    author         = "{ALICE Collaboration}",
      $author = "Acharya, Shreyasi and others",
    collaboration = "ALICE",
    title = "{Energy dependence of forward-rapidity $\mathrm {J}/\psi $ and $\psi \mathrm {(2S)}$ production in pp collisions at the LHC}",
    eprint = "1702.00557",
    archivePrefix = "arXiv",
    primaryClass = "hep-ex",
    reportNumber = "CERN-EP-2017-015",
    doi = "10.1140/epjc/s10052-017-4940-4",
    journal = "Eur. Phys. J. C",
    volume = "77",
    number = "6",
    pages = "392",
    year = "2017"
}

@article{ALICE5,
    author         = "{ALICE Collaboration}",
    collaboration = "ALICE",
    title = "{Measurement of prompt $J/\psi$ and beauty hadron production cross sections at mid-rapidity in $pp$ collisions at $\sqrt{s} = 7$ TeV}",
    eprint = "1205.5880",
    archivePrefix = "arXiv",
    primaryClass = "hep-ex",
    reportNumber = "CERN-PH-EP-2012-132",
    doi = "10.1007/JHEP11(2012)065",
    journal = "JHEP",
    volume = "11",
    pages = "065",
    year = "2012"
}

@techreport{ATL-SOFT-PUB-2021-001,
      author         = "{ATLAS Collaboration}",
      collaboration = "ATLAS",
      title         = "{The ATLAS Collaboration Software and Firmware}",
      institution   = "CERN",
      reportNumber  = "ATL-SOFT-PUB-2021-001",
      address       = "Geneva",
      year          = "2021",
      url           = "https://cds.cern.ch/record/2767187",
      %note          = "https://atlas.web.cern.ch/Atlas/GROUPS/PHYSICS/PUBNOTES/ATL-SOFT-PUB-2021-001",
}

@inproceedings{Jung:2004gs,
    author = "Jung, Hannes",
    title = "{Un-integrated PDFs in CCFM}",
    booktitle = "{12th International Workshop on Deep Inelastic Scattering (DIS 2004)}",
    doi = "10.48550/arXiv.hep-ph/0411287",
    eprint = "hep-ph/0411287",
    archivePrefix = "arXiv",
    pages = "299--302",
    month = "11",
    year = "2004"
}

@Article{PERF-2015-10,
    author         = "{ATLAS Collaboration}",
    title          = "{Muon reconstruction performance of the ATLAS detector in proton--proton collision data at \(\sqrt{s} = 13\,\text{TeV}\)}",
    journal        = "Eur. Phys. J. C",
    volume         = "76",
    year           = "2016",
    pages          = "292",
    doi            = "10.1140/epjc/s10052-016-4120-y",
    reportNumber   = "CERN-EP-2016-033",
    eprint         = "1603.05598",
    archivePrefix  = "arXiv",
    primaryClass   = "hep-ex",
}

@article{Sjostrand:2014zea,
    author = {Sj\"ostrand, Torbj\"orn and Ask, Stefan and Christiansen, Jesper R. and Corke, Richard and Desai, Nishita and Ilten, Philip and Mrenna, Stephen and Prestel, Stefan and Rasmussen, Christine O. and Skands, Peter Z.},
    title = "{An introduction to PYTHIA 8.2}",
    eprint = "1410.3012",
    archivePrefix = "arXiv",
    primaryClass = "hep-ph",
    reportNumber = "LU-TP-14-36, MCNET-14-22, CERN-PH-TH-2014-190, FERMILAB-PUB-14-316-CD, DESY-14-178, SLAC-PUB-16122",
    doi = "10.1016/j.cpc.2015.01.024",
    journal = "Comput. Phys. Commun.",
    volume = "191",
    pages = "159--177",
    year = "2015"
}

@techreport{ATL-PHYS-PUB-2014-021,
      author         = "{ATLAS Collaboration}",
      title         = "{ATLAS Pythia 8 tunes to 7 TeV data}",
      institution   = "CERN",
      reportNumber  = "ATL-PHYS-PUB-2014-021",
      address       = "Geneva",
      year          = "2014",
      url           = "https://cds.cern.ch/record/1966419",
      %note          = "All figures including auxiliary figures are available at
        %https://atlas.web.cern.ch/Atlas/GROUPS/PHYSICS/PUBNOTES/ATL-PHYS-PUB-2014-021",
}

@article{DAPR-2021-01,
    author         = "{ATLAS Collaboration}",
    collaboration = "ATLAS",
    title = "{Luminosity determination in $pp$ collisions at $\sqrt{s}=13$ TeV using the ATLAS detector at the LHC}",
    eprint = "2212.09379",
    archivePrefix = "arXiv",
    primaryClass = "hep-ex",
    reportNumber = "CERN-EP-2022-281",
    doi = "10.1140/epjc/s10052-023-11747-w",
    journal = "Eur. Phys. J. C",
    volume = "83",
    number = "10",
    pages = "982",
    year = "2023"
}

@article{ALICE:2011gej,
    author         = "{ALICE Collaboration}",
      %author = "Abelev, Betty and others",
    collaboration = "ALICE",
    title = "{$J/\psi$ polarization in $pp$ collisions at $\sqrt{s}=7$ TeV}",
    eprint = "1111.1630",
    archivePrefix = "arXiv",
    primaryClass = "hep-ex",
    reportNumber = "CERN-PH-EP-2011-182",
    doi = "10.1103/PhysRevLett.108.082001",
    journal = "Phys. Rev. Lett.",
    volume = "108",
    pages = "082001",
    year = "2012"
}

@article{Ma:2018qvc,
    author = "Ma, Yan-Qing and Stebel, Tomasz and Venugopalan, Raju",
    title = "{$J/\psi$ polarization in the CGC+NRQCD approach}",
    eprint = "1809.03573",
    archivePrefix = "arXiv",
    primaryClass = "hep-ph",
    doi = "10.1007/JHEP12(2018)057",
    journal = "JHEP",
    volume = "12",
    pages = "057",
    year = "2018"
}

@article{Feng:2018ukp,
    author = "Feng, Yu and Gong, Bin and Chang, Chao-Hsi and Wang, Jian-Xiong",
    title = "{Remaining parts of the long-standing $J/\psi$ polarization puzzle}",
    eprint = "1810.08989",
    archivePrefix = "arXiv",
    primaryClass = "hep-ph",
    doi = "10.1103/PhysRevD.99.014044",
    journal = "Phys. Rev. D",
    volume = "99",
    number = "1",
    pages = "014044",
    year = "2019"
}

@article{Faccioli:2022alj,
    author = "Faccioli, Pietro and Louren\c{c}o, Carlos",
    title = "{On the polarization of the non-prompt contribution to inclusive J/\ensuremath{\psi} production in pp collisions}",
    eprint = "2206.14686",
    archivePrefix = "arXiv",
    primaryClass = "hep-ph",
    doi = "10.1007/JHEP10(2022)010",
    journal = "JHEP",
    volume = "10",
    pages = "005",
    year = "2022"
}

@article{PhysRevLett.85.2886,
  author         = "{CDF Collaboration}",
  title = {Measurement of $\mathit{J}/\ensuremath{\psi}$ and $\ensuremath{\psi}(2\mathit{S})$ Polarization in $\mathit{p}\overline{p}$ Collisions at $\ensuremath{\surd}\mathit{s}\phantom{\rule{0ex}{0ex}}=\phantom{\rule{0ex}{0ex}}1.8\mathrm{TeV}$},
  collaboration = {CDF Collaboration},
  journal = {Phys. Rev. Lett.},
  volume = {85},
  pages = {2886--2891},
  numpages = {0},
  year = {2000},
  month = {Oct},
  publisher = {American Physical Society},
  doi = {10.1103/PhysRevLett.85.2886},
}

@phdthesis{CB,
    author = "Oreglia, M.",
    title = "{A Study of the Reactions $\psi^\prime \to \gamma \gamma \psi$}",
    school  = "",
    reportNumber = "SLAC-0236, SLAC-236, UMI-81-08973, SLAC-R-0236, SLAC-R-236",
    type = "PhD thesis, Stanford University",
    year = "1980",
    addendum       = " \href{https://www-public.slac.stanford.edu/sciDoc/docMeta.aspx?slacPubNumber=slac-r-236}{SLAC-R-236}",      
}

@article{Barger:1979js,
    author = "Barger, Vernon D. and Keung, Wai-Yee and Phillips, R. J. N.",
    title = "{On psi and Upsilon Production via Gluons}",
    reportNumber = "Print-79-0776 (WISCONSIN)",
    doi = "10.1016/0370-2693(80)90444-X",
    journal = "Phys. Lett. B",
    volume = "91",
    pages = "253--258",
    year = "1980"
}

@article{Barger:1980mg,
    author = "Barger, Vernon D. and Keung, Wai-Yee and Phillips, R. J. N.",
    title = "{Hadroproduction of $\psi$ and $\Upsilon$}",
    reportNumber = "COO-881-139",
    doi = "10.1007/BF01588844",
    journal = "Z. Phys. C",
    volume = "6",
    pages = "169",
    year = "1980"
}

@article{Gavai:1994in,
    author = "Gavai, R. and Kharzeev, D. and Satz, H. and Schuler, G. A. and Sridhar, K. and Vogt, R.",
    title = "{Quarkonium production in hadronic collisions}",
    eprint = "hep-ph/9502270",
    archivePrefix = "arXiv",
    reportNumber = "CERN-TH-7526-94, BI-TP-63-94, CERN-TH.7526-94",
    doi = "10.1142/S0217751X95001443",
    journal = "Int. J. Mod. Phys. A",
    volume = "10",
    pages = "3043--3070",
    year = "1995"
}

@article{Ma:2016exq,
    author = "Ma, Yan-Qing and Vogt, Ramona",
    title = "{Quarkonium Production in an Improved Color Evaporation Model}",
    eprint = "1609.06042",
    archivePrefix = "arXiv",
    primaryClass = "hep-ph",
    doi = "10.1103/PhysRevD.94.114029",
    journal = "Phys. Rev. D",
    volume = "94",
    number = "11",
    pages = "114029",
    year = "2016"
}

@article{Agostinelli:2002hh,
      author         = "Agostinelli, S. and others",
      title          = "{\textsc{Geant4} -- a simulation toolkit}",
      journal        = "Nucl. Instrum. Meth. A",
      volume         = "506",
      year           = "2003",
      pages          = "250",
      doi            = "10.1016/S0168-9002(03)01368-8",
      reportNumber   = "SLAC-PUB-9350, FERMILAB-PUB-03-339",
      SLACcitation   = "%%CITATION = NUIMA,A506,250;%%"
}

@article{Zyla:2020zbs,
    author = "{Particle Data Group} and Zyla, P.A. and others",
    title = "{Review of Particle Physics}",
    journal = "PTEP",
    volume = "2020",
    number = "8",
    pages = "083C01",
    year = "2020",
    doi = "10.1093/ptep/ptaa104",
}

@Article{PERF-2007-01,
    author         = "{ATLAS Collaboration}",
    title          = "{The ATLAS Experiment at the CERN Large Hadron Collider}",
    journal        = "JINST",
    volume         = "3",
    year           = "2008",
    pages          = "S08003",
    doi            = "10.1088/1748-0221/3/08/S08003",
    primaryClass   = "hep-ex",
}

@Article{SOFT-2010-01,
    author         = "{ATLAS Collaboration}",
    title          = "{The ATLAS Simulation Infrastructure}",
    journal        = "Eur. Phys. J. C",
    volume         = "70",
    year           = "2010",
    pages          = "823",
    doi            = "10.1140/epjc/s10052-010-1429-9",
    eprint         = "1005.4568",
    archivePrefix  = "arXiv",
    primaryClass   = "physics.ins-det",
}

@Article{BPHY-2012-02,
    author         = "{ATLAS Collaboration}",
    title          = "{Measurement of the differential cross-sections of prompt and non-prompt production of \(J/\psi\) and \(\psi(2\text{S})\) in \(pp\) collisions at \(\sqrt{s} = 7\) and \(8\,\text{TeV}\) with the ATLAS detector}",
    journal        = "Eur. Phys. J. C",
    volume         = "76",
    year           = "2016",
    pages          = "283",
    doi            = "10.1140/epjc/s10052-016-4050-8",
    reportNumber   = "CERN-PH-EP-2015-292",
    eprint         = "1512.03657",
    archivePrefix  = "arXiv",
    primaryClass   = "hep-ex",
}

@Article{BPHY-2013-05,
    author         = "{ATLAS Collaboration}",
    title          = "{Measurement of \(\chi_{c1}\) and \(\chi_{c2}\) production with \(\sqrt{s} = 7\,\text{TeV}\) \(pp\) collisions at ATLAS}",
    journal        = "JHEP",
    volume         = "07",
    year           = "2014",
    pages          = "154",
    doi            = "10.1007/JHEP07(2014)154",
    reportNumber   = "CERN-PH-EP-2013-202",
    eprint         = "1404.7035",
    archivePrefix  = "arXiv",
    primaryClass   = "hep-ex",
}

@Article{BPHY-2013-06,
    author         = "{ATLAS Collaboration}",
    title          = "{Measurement of the production cross-section of \(\psi(2\text{S})\rightarrow J/\psi(\rightarrow\mu^+\mu^-)\pi^+\pi^-\) in \(pp\) collisions at \(\sqrt{s} = 7\,\text{TeV}\) at ATLAS}",
    journal        = "JHEP",
    volume         = "09",
    year           = "2014",
    pages          = "079",
    doi            = "10.1007/JHEP09(2014)079",
    reportNumber   = "CERN-PH-EP-2014-095",
    eprint         = "1407.5532",
    archivePrefix  = "arXiv",
    primaryClass   = "hep-ex",
}

@Article{BPHY-2019-08,
    author         = "{ATLAS Collaboration}",
    title          = "{Measurement of the production cross-section of \(J/\psi\) and \(\psi(2\text{S})\) mesons in \(pp\) collisions at \(\sqrt{s} = 13\,\text{TeV}\) with the ATLAS detector}",
    year           = "2023",
    reportNumber   = "CERN-EP-2023-193",
    eprint         = "2309.17177",
    archivePrefix  = "arXiv",
    primaryClass   = "hep-ex",
}

@Booklet{ATL-SOFT-PUB-2023-001,
    author         = "{ATLAS Collaboration}",
    title          = "{ATLAS Computing Acknowledgements}",
    howpublished   = "{ATL-SOFT-PUB-2023-001}",
    url            = "https://cds.cern.ch/record/2869272",
    year           = "2023",
}

@Article{CMS-BPH-13-003,
    author         = "{CMS Collaboration}",
    title          = "{Measurement of the prompt \(J/\psi\) and \(\psi(2\text{S})\) polarizations in \(pp\) collisions at \(\sqrt{s} = 7\,\text{TeV}\)}",
    journal        = "Phys. Lett. B",
    volume         = "727",
    year           = "2013",
    pages          = "381",
    doi            = "10.1016/j.physletb.2013.10.055",
    reportNumber   = "CERN-PH-EP-2013-140",
    eprint         = "1307.6070",
    archivePrefix  = "arXiv",
    primaryClass   = "hep-ex",
}

@Article{CMS-BPH-14-001,
    author         = "{CMS Collaboration}",
    title          = "{Measurement of \(J/\psi\) and \(\psi(2\text{S})\) Prompt Double-Differential Cross Sections in \(pp\) Collisions at \(\sqrt{s} = 7\,\text{TeV}\)}",
    journal        = "Phys. Rev. Lett.",
    volume         = "114",
    year           = "2015",
    pages          = "191802",
    doi            = "10.1103/PhysRevLett.114.191802",
    reportNumber   = "CERN-PH-EP-2015-022",
    eprint         = "1502.04155",
    archivePrefix  = "arXiv",
    primaryClass   = "hep-ex",
}

@Article{CMS-BPH-15-005,
    author         = "{CMS Collaboration}",
    title          = "{Measurement of quarkonium production cross sections in \(pp\) collisions at \(\sqrt{s} = 13\,\text{TeV}\)}",
    journal        = "Phys. Lett. B",
    volume         = "780",
    year           = "2018",
    pages          = "251",
    doi            = "10.1016/j.physletb.2018.02.033",
    reportNumber   = "CERN-EP-2017-267",
    eprint         = "1710.11002",
    archivePrefix  = "arXiv",
    primaryClass   = "hep-ex",
}
 
\clearpage
% ATLAS Collaboration author list
% Reference date of BPHY-2019-08 is 2023-06-16
% Author list last updated on date 22-FEB-24
% Data extracted on 22-Feb-2024 for paper reference BPHY-2019-08
% at 4:37pm
 
\begin{flushleft}
\hypersetup{urlcolor=black}
{\Large The ATLAS Collaboration}

\bigskip

\AtlasOrcid[0000-0002-6665-4934]{G.~Aad}$^\textrm{\scriptsize 102}$,
\AtlasOrcid[0000-0002-5888-2734]{B.~Abbott}$^\textrm{\scriptsize 120}$,
\AtlasOrcid[0000-0002-1002-1652]{K.~Abeling}$^\textrm{\scriptsize 55}$,
\AtlasOrcid[0000-0001-5763-2760]{N.J.~Abicht}$^\textrm{\scriptsize 49}$,
\AtlasOrcid[0000-0002-8496-9294]{S.H.~Abidi}$^\textrm{\scriptsize 29}$,
\AtlasOrcid[0000-0002-9987-2292]{A.~Aboulhorma}$^\textrm{\scriptsize 35e}$,
\AtlasOrcid[0000-0001-5329-6640]{H.~Abramowicz}$^\textrm{\scriptsize 151}$,
\AtlasOrcid[0000-0002-1599-2896]{H.~Abreu}$^\textrm{\scriptsize 150}$,
\AtlasOrcid[0000-0003-0403-3697]{Y.~Abulaiti}$^\textrm{\scriptsize 117}$,
\AtlasOrcid[0000-0002-8588-9157]{B.S.~Acharya}$^\textrm{\scriptsize 69a,69b,m}$
% ...
\bigskip
\\

% ...
$^{29}$Physics Department, Brookhaven National Laboratory, Upton NY; United States of America.\\
% ...
$^{35}$$^{(a)}$Facult\'e des Sciences Ain Chock, R\'eseau Universitaire de Physique des Hautes Energies - Universit\'e Hassan II, Casablanca;$^{(b)}$Facult\'{e} des Sciences, Universit\'{e} Ibn-Tofail, K\'{e}nitra;$^{(c)}$Facult\'e des Sciences Semlalia, Universit\'e Cadi Ayyad, LPHEA-Marrakech;$^{(d)}$LPMR, Facult\'e des Sciences, Universit\'e Mohamed Premier, Oujda;$^{(e)}$Facult\'e des sciences, Universit\'e Mohammed V, Rabat;$^{(f)}$Institute of Applied Physics, Mohammed VI Polytechnic University, Ben Guerir; Morocco.\\
% ...
$^{49}$Fakult\"{a}t Physik , Technische Universit{\"a}t Dortmund, Dortmund; Germany.\\
% ...
$^{55}$II. Physikalisches Institut, Georg-August-Universit\"{a}t G\"ottingen, G\"ottingen; Germany.\\
% ...
$^{69}$$^{(a)}$INFN Gruppo Collegato di Udine, Sezione di Trieste, Udine;$^{(b)}$ICTP, Trieste;$^{(c)}$Dipartimento Politecnico di Ingegneria e Architettura, Universit\`a di Udine, Udine; Italy.\\
% ...
$^{102}$CPPM, Aix-Marseille Universit\'e, CNRS/IN2P3, Marseille; France.\\
% ...
$^{117}$Department of Physics, New York University, New York NY; United States of America.\\
% ...
$^{120}$Homer L. Dodge Department of Physics and Astronomy, University of Oklahoma, Norman OK; United States of America.\\
% ...
$^{150}$Department of Physics, Technion, Israel Institute of Technology, Haifa; Israel.\\
$^{151}$Raymond and Beverly Sackler School of Physics and Astronomy, Tel Aviv University, Tel Aviv; Israel.\\
% ...
$^{a}$ Also Affiliated with an institute covered by a cooperation agreement with CERN.\\
$^{b}$ Also at An-Najah National University, Nablus; Palestine.\\
$^{c}$ Also at Borough of Manhattan Community College, City University of New York, New York NY; United States of America.\\
$^{d}$ Also at Center for High Energy Physics, Peking University; China.\\
$^{e}$ Also at Center for Interdisciplinary Research and Innovation (CIRI-AUTH), Thessaloniki; Greece.\\
$^{f}$ Also at Centro Studi e Ricerche Enrico Fermi; Italy.\\
% ...
$^{m}$ Also at Department of Physics, King's College London, London; United Kingdom.\\
% ...

\end{flushleft}

% Created with Glance <Atlas.Glance@cern.ch>

\end{document}